\documentclass{natureprintstyle}
\usepackage[sort&compress,numbers]{natbib}
\usepackage[utf8x]{inputenc}
\usepackage{amsfonts,amsmath,amssymb,stmaryrd}

\usepackage{graphicx}
\usepackage{siunitx} 
\usepackage{braket}
\usepackage{multirow}
\usepackage[super]{nth}
\graphicspath{{./images/figures/}}
\usepackage{siunitx}
\usepackage{changes}
\usepackage[hidelinks]{hyperref}
\usepackage[nameinlink]{cleveref}
\Crefname{equation}{Eq.}{Eqs.}
\Crefname{figure}{Fig.}{Figs.}
\Crefname{tabular}{Tab.}{Tabs.}
\Crefname{table}{Tab.}{Tabs.}

\usepackage{nicefrac}
\usepackage{tikz}
\usepackage{tabularx}
\usepackage{bm} 
\usepackage{color}

\DeclareSIUnit\gauss{G}
\DeclareSIUnit\bohrradii{a_0}

\title{Indication of critical scaling in time during the relaxation of an open quantum system}

  \author{Ling-Na Wu$^{1,2,4}$, Jens Nettersheim$^{3,4}$, Julian Fe\ss$^3$, Alexander Schnell$^1$, Sabrina Burgardt$^3$, Silvia Hiebel$^3$, Daniel Adam$^3$, Andr{\'e} Eckardt$^{1,5}$, and Artur Widera$^{3,6}$}

\begin{document}

\maketitle

\begin{affiliations}
\item Institut f\"ur Theoretische Physik, Technische Universit\"at Berlin,
Hardenbergstra\ss e 36, 10623 Berlin, Germany
 \item Center for Theoretical Physics and School of Science, Hainan University, Haikou 570228, China
\item Department of Physics and State Research Center OPTIMAS, Technische Universit\"at Kaiserslautern, 67663 Kaiserslautern, Germany
\item These authors contributed equally to this work: Ling-Na Wu and Jens Nettersheim.
\item email: eckardt@tu-berlin.de
\item email: widera@physik.uni-kl.de
\end{affiliations}

		\date{\today}


	\maketitle
\begin{abstract}
Phase transitions correspond to the singular behavior of physical systems in response to continuous control parameters like temperature or external fields~\cite{domb2000phase}. 
Near continuous phase transitions, associated with the divergence of a correlation length, universal power-law scaling behavior with critical exponents independent of microscopic system details is found.
Recently, dynamical quantum phase transitions and universal scaling have been predicted and also observed in the non-equilibrium dynamics of isolated quantum systems after a quench, with time playing the role of the control parameter~\cite{Berges2008,Prufer2018,Erne2018,Eigen2018,Schmied2019,Heyl2013PRL,Jurcevic2017, Flaeschner2018,PhysRevLett.122.020501,Heyl2018}.
However, signatures of such critical phenomena in time in open systems, whose dynamics is driven by the dissipative contact to an environment, were so far elusive.
Here, we present {results indicating} that critical scaling with respect to time can also occur {during the relaxation dynamics of an} open quantum system described by mixed states. We experimentally measure the relaxation dynamics of the large atomic spin of individual Caesium atoms induced by the dissipative coupling via spin-exchange processes to an ultracold Bose gas of Rubidium atoms. For initial states far from equilibrium, the entropy of the spin state is found to peak in time, transiently approaching its maximum possible value, before eventually relaxing to its lower equilibrium value. Moreover, a finite-size scaling analysis based on numerical simulations shows that it corresponds to a critical point with respect to time of the dissipative system in the limit of large system sizes. {It is signalled by the divergence of a characteristic length at a critical time, characterized by critical exponents that are found to be independent of system details.}

\end{abstract}	

 
Phase transitions emerge from the collective behavior of large quantum systems in the thermodynamic limit \cite{domb2000phase}. 
A continuous phase transition is signaled by the divergence of a characteristic length scale $\xi$, when the control parameter approaches a critical value. As a result, the behavior near the transition becomes independent of the microscopic details of a system, giving rise to universal critical exponents \cite{domb2000phase}, like the one describing the divergence of $\xi$ as a function of the control parameter. 
Despite the fact that the distinction between different phases of matter, like liquid or crystalline, is an essential and well-known aspect of nature, phase transitions and their critical behavior remain an active field of research until today. Subjects of interest include, for instance,
quantum phase transitions happening in pure quantum ground states at absolute zero~\cite{sachdev_2011} and
topological phase transitions beyond Landau's paradigm~\cite{Wen17}. 

	\begin{figure}[!htp]    
    \includegraphics[width=0.99\columnwidth]{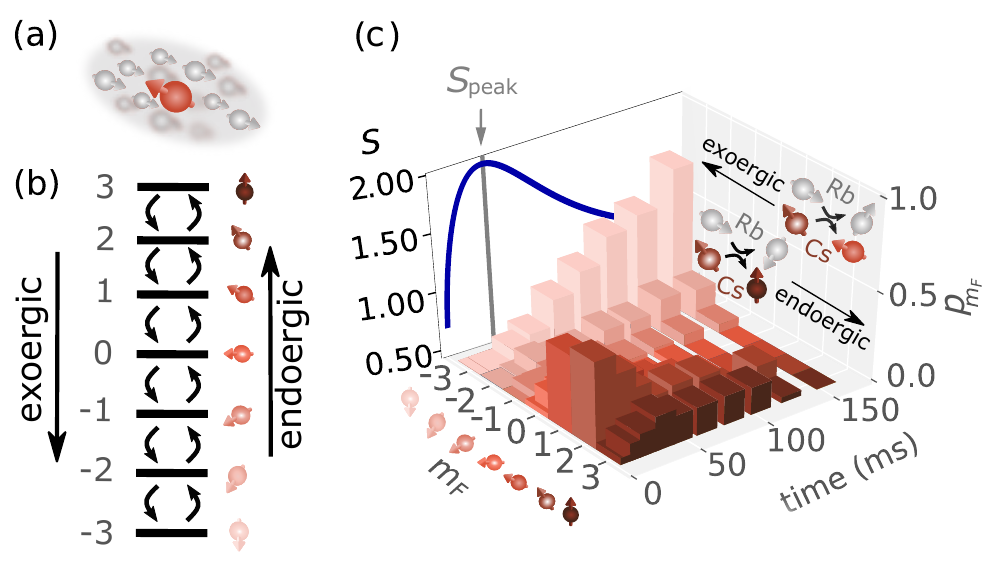}
	\caption{{\bf Realizing an open spin system} (a) Individual Cs atom (red) interacting with a bath of spin-polarized Rb atoms via inelastic spin-exchange~(SE) collisions.
	(b) The Cs-Zeeman states experimentally realize an equidistant seven-level ($m_{F} \in 3, 2, ..., -3$) spin system defining $m_{F}=-3$ as ground state for the Rb-Cs compound (for more details, see Methods). SE collisions with the Rb atoms give rise to dissipative spin dynamics, increasing (decreasing) internal energy and angular momentum for endoergic (exoergic) processes.
    The twelve SE rates between the Cs Zeeman states depend on the external magnetic field and the bath temperature. (c) Bath-driven and time-resolved quantum-spin evolution for individual Cs atoms initially prepared in a mixture of $\ket{m_F=1}$ and $\ket{m_F=2}$.
    Sketches in the back panel show the microscopic collision processes of exoergic  
    and endoergic SE collisions. The lateral plane shows the resulting entropy evolution, featuring a maximum at $S_{\rm peak}= 1.944 \approx \ln 7 =S_{\rm max}$.}
	\label{models2}
    \end{figure}
    

Recently, the transient evolution of isolated quantum systems gained considerable interest, as it can be realized in engineered quantum systems such as ultracold atomic quantum gases. Prominent effects that were studied include the transition between eigenstate thermalization and many-body localization~\cite{Nandkishore2015MBLReview,RevModPhys.83.863,choi2016exploring}, non-equilibrium phase transitions in the long-time (or prethermal) behavior of (almost) integrable quantum systems~\cite{Albiez2005, Marino2022}, or the observation of discrete time crystals in interacting Floquet systems~\cite{Khemani2016,Else2016}. Another fascinating example is the prediction and observation of dynamical quantum phase transitions~\cite{Heyl2013PRL,Jurcevic2017, Flaeschner2018,PhysRevLett.122.020501,Heyl2018} and universal scaling behavior~\cite{Berges2008, Prufer2018,Erne2018,Eigen2018,Schmied2019} occurring at a critical time during the transient non-equilibrium evolution of isolated quantum systems.  Here \emph{time} plays the role of the control parameter. The underlying non-equilibrium dynamics can be initialized, for example, by a quantum quench, i.e.,\ a rapid parameter variation starting from the ground state of the previous Hamiltonian.
%

 \begin{figure*}[ht] 
        \includegraphics[width=2\columnwidth]{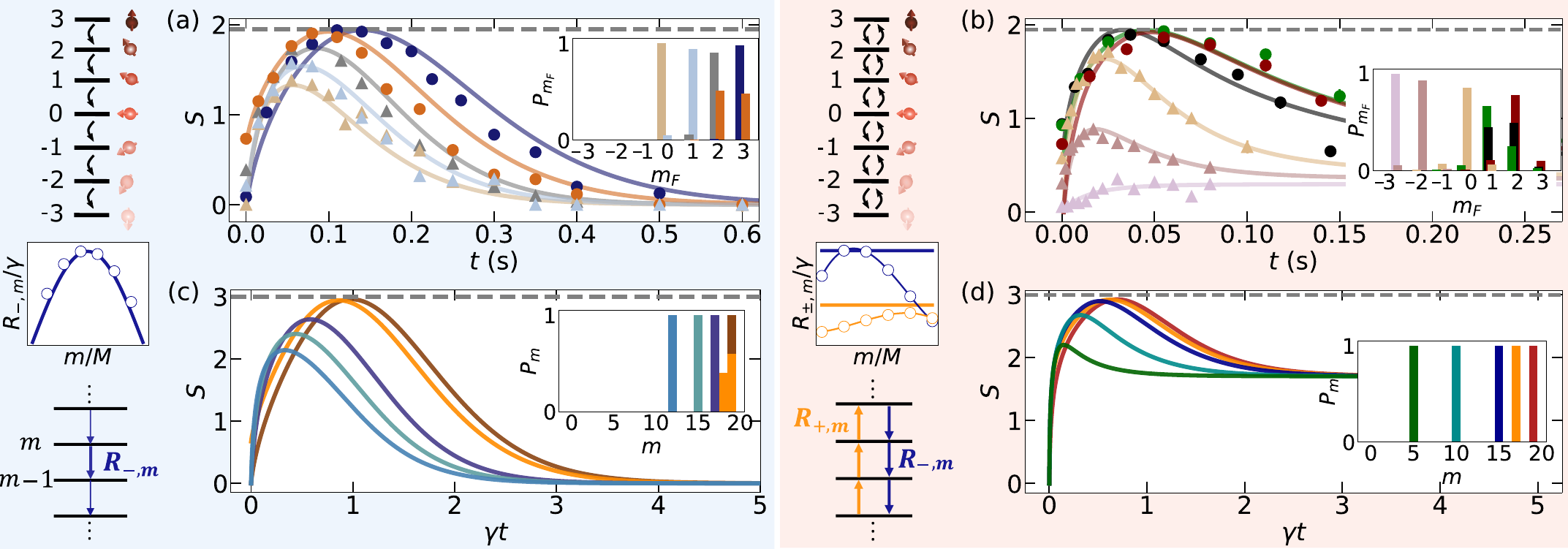}
	\caption{{\bf Entropy evolution} Blue (red) background indicates regimes where unidirectional spin transitions, applying a high magnetic field ($B=460$~mG)  and bidirectional spin transitions, utilizing a low magnetic field ($B=25$~mG) are possible. While in the former case, endoergic transitions are suppressed completely~\cite{Bouton_Spin_Thermometer}, in the latter case, they are allowed and raise $m_{F}$ for Cs, but with reduced probability compared to exoergic processes, which lower $m_{F}$.
	(a, b) Experimentally measured entropy evolution starting from different initial states shown in the insets (colors match). Bullets ($\bullet$) are used for trajectories with high peak entropy ($S_{\rm peak}\ge 0.98 S_{\rm max}$) and triangles ($\blacktriangle$) otherwise. Error bars are smaller than symbol sizes and represent the standard deviation. 
	Solid lines are obtained from simulations, dashed lines indicate maximum possible entropy $S_{\rm max}$. 
	(c, d) Like (a, b) but for theoretical models with 20 states. Panels between the level schemes show normalized transfer rates for the experimental system (dots) and the theoretical models (lines). For the unidirectional model~(in blue background), $R_{-,m} = \gamma M \sin(\pi (m+1)/M)$, where $\gamma$ is the coupling strength. For the bidirectional model~(red background), thick horizontal lines correspond to the state-independent-rate model, with  $R_{+,m} \equiv R_{+}=\gamma M$~(orange) and $R_{-,m} \equiv R_{-}=\gamma M \exp(10/M)$~(blue). 
 }
	\label{entropy_evolution}
    \end{figure*}
    
%
%

{In the following, we describe {another} example of critical behavior with respect to time, reminiscent of a continuous phase transition {associated with the dynamics of a quantum system.} {It happens during the relaxation of} an \emph{open} system {and} corresponds to the divergence of a localization length $\xi$ at a critical time. It is, thus, different from the dynamical quantum phase transitions associated with the non-analytic behavior of the return probability in isolated systems, as they were described previously \cite{Heyl2018}. Importantly, it is also different compared to non-equilibrium phase transitions occurring in the long-time behaviour of
driven-dissipative quantum systems in response to control parameters other than time (see, e.g., Ref.~\cite{sieberer2016keldysh}). } 

In contrast to isolated quantum systems, open quantum systems~\cite{breuer2002theory} are characterized by the coupling to an environment, called a bath, with which they exchange both energy and information. Markovian baths rapidly dissipate information, so that the dynamics of the system can be described by an idealized time-local master equation $\dot\rho(t)=\mathcal{L}[\rho(t)]$, where the dynamics is generated by a time-independent Liouvillian superoperator $\mathcal{L}$ acting on the instantaneous density operator $\rho(t)$ describing the system's state at time $t$. If the coupling to the environment is weak compared to the level spacing in the system, $\rho(t)$ quickly becomes diagonal with respect to the energy eigenstates $|m\rangle$, $\rho(t)\simeq \sum_m p_m(t)|m\rangle\langle m|$. The probabilities $p_m(t)$ for being in state $|m\rangle$ then follow a Pauli rate equation $\dot{p}_m=\sum_{m'\ne m}[R_{mm'} p_{m'} -R_{m'm}p_m]$, with $R_{mm'}$ denoting the rate for a bath-induced transition from $|m'\rangle$ to~$|m\rangle$~\cite{breuer2002theory}. 

In systems of ultracold atoms, dissipation can be engineered in various ways, including, for instance, the coupling of the atoms to a cavity \cite{Ritsch2013}, spontaneous emission of lattice photons \cite{Pichler13,Lueschen17}, particle loss (e.g.\ via controlled ionization \cite{Brazhnyi09}), or the coupling to a background gas \cite{Diehl08}.
We realize such an open system by the spin degrees of freedom of individual ultracold Caesium atoms ($^{133}$Cs), which are immersed as impurities in a bath comprising ultracold Rubidium atoms ($^{87}$Rb) [see the sketch in Fig.~\ref{models2}(a) and Methods for details]. 
The hyperfine states of both species form stable quasi-spins with quantum numbers $F=3$ ($F=1$) for Cs (Rb). 
In the presence of a weak, constant external magnetic field $B$, the spins possess an equidistant ladder spectrum $E_{m_F} = m_F\Delta$, where $\Delta= g_\mathrm{F} \mu_\mathrm{B} B / \hbar$, with Landé factor $g_\mathrm{F}$, reduced Planck constant $\hbar$ and Bohr magneton $\mu_\mathrm{B}$.  
The corresponding energy eigenstates $|m_F\rangle$ are characterized by the magnetic quantum number $m_F=-F,-F+1,\ldots,F$. 
{While the Cs $\vert m_F=3\rangle$ state is the ground state of the isolated Cs atom, it is the highest excited Cs state of the open Cs-Rb system. Controlling the initial Cs-state population allows experimentally initializing the open-system dynamics with almost arbitrary excitation energy of the Cs spin.}
Elastic Rb-Cs collisions quickly thermalize the Cs atoms' center-of-mass motion, while inelastic spin-exchange (SE) processes give rise to bath-induced transitions, where the Cs spins are changed by single quanta of angular momentum, $m_{F}\to m'_{F} =m_{F}\pm 1$ with corresponding rates $R_{\pm,m_F} \equiv R_{m_F\pm1,m_F}$~\cite{Schmidt_Tailored}, see Fig.~\ref{models2}(b) and Methods ($m_F$ is used throughout for the Cs spins). 
\textcolor{black}{The combination of a large atom-number imbalance, the ratio of elastic to inelastic collision rates, and a relatively large mean-free path realize an almost ideal Markov bath, yielding a collision probability of a Cs impurity with the same Rb atom of well below a percent (see supplemental material).}
We initially prepare the Cs impurity in an excited spin state
defined by the probability distribution $p_{m_F}(0)$, and monitor the subsequent relaxation dynamics $p_{m_F}(t)$; see Fig.~\ref{models2}(c) for an example. 
As an important observable, we extract the evolution of the total entropy of our spin system [blue curve in Fig.~\ref{models2}(c)],
\begin{eqnarray}
\displaystyle S(t) = - \sum_{m_{F}} p_{m_{F}}(t) \ln(p_{m_{F}}(t)).
\end{eqnarray}

In Fig.~\ref{entropy_evolution}(a) and (b), we show the measured evolution of $S$ for various different initial conditions (specified in the insets). This papers's blue and red background colors indicate unidirectional and bidirectional spin-exchange.
For highly excited initial states, that means for the Rb-Cs compound states of large positive $m_F$ (for more details, see Methods), we find in both scenarios that the entropy evolution is highly non-monotonous. The entropy first increases to reach a peak value $S_{\rm peak}$ at a time $t_{\rm peak}$, before eventually relaxing to a steady-state. This is remarkable and rather different from the behavior found for initial states close to equilibrium, for which we observe that the entropy simply increases in time until it saturates at its steady-state value [see pink curve in Fig.~\ref{entropy_evolution}(b)].
Even more remarkably, for various different initial conditions\textcolor{black}{, i.e., different $m_F$-states and their combinations}\footnote{\textcolor{black}{See supplemental material for a discussion of the role of the initial state and its energy.}}, this peak value almost reaches the maximal possible entropy, $S_{\rm max}=\ln{M}\approx1.95$ with $M=7$ being the number of spin states, indicated by the dashed line.

\textcolor{black}{Experimentally, the total signal of $m_F$ populations as well as of the entropy value is the average over Cs-atom signals locally interacting with the inhomogeneous density distribution of the bath.} {Importantly, since the Cs impurities undergo approximately ten elastic collisions between two spin-exchange collisions and, moreover, the Cs mean-free path in the Rb cloud is of the order of the Rb cloud's extension (see supplemental material), each Cs atom samples the whole inhomogeneous density during its relaxation. Therefore, the spin-population dynamics reflects a homogeneous broadening rather than local dynamics.}
Furthermore, since the number of $m_F$  states is bound for the Cs impurity, an average can only lower the total entropy signal, so that the measured values close to the maximum value are a lower bound. 

The approach of $S_{\rm max}$ implies that the system transiently approaches the maximally mixed state $\rho_{\rm max}=M^{-1}\sum_{m_F}|m_F\rangle\langle m_F|$, corresponding to a completely delocalized spin distribution $p_{m_F}=1/M$. In the limit of large $M$, such behavior implies a divergence of both $S$ and the ``length'' 
$\xi$ {that characterizes the number of states $\ket{m_F}$ covered by the probability distribution $p_{m_F}$. The latter can, e.g., be defined as the participation ratio, $\xi\equiv(\sum_{m_F} p_{m_F}^2)^{-1}$}. 
This, in turn, directly corresponds to the divergence of a relevant length scale $\xi$ that is found when a system approaches a critical point like at a continuous phase transition. Here, however, the continuous control parameter is the time $t$ and its critical value is $t_{\rm peak}$. In this sense, a transient approach of the maximally mixed state $\rho_{\rm max}$, indicates a phase transition {in time} in the limit of large~$M$. 


To answer the question whether the observed dynamics is indeed a finite-size precursor of {a} phase transition, we define two model systems of variable system size $M$~[see Fig.~\ref{entropy_evolution} middle and lower side panels]
and numerically perform a finite-size scaling analysis to extract the behavior for $M\to\infty$. Both models consist of $M$ states labeled by $m=0,1,\dots ,M-1$, which form an equidistant energy spectrum $\varepsilon_m=m\Delta$.  
A unidirectional model generalizes the high-magnetic-field regime to larger $M$. Here, only transitions $|m\rangle \to |m'=m-1\rangle$ occur, corresponding to a zero-temperature bath. The rates $R_{-,m}$ possess a parabolic dependence on $m$ mimicking the experimental rates. {Such a rate inhomogeneity is required for reaching high peak entropies, since for unidirectional transport a right-moving probability distribution can only become broader if the velocity at its right end is larger than at its left end. }
In a bidirectional model, corresponding to the case of low magnetic fields, it is sufficient to assume state independent rates.
Figs.~\ref{entropy_evolution}(c) and (d) depict the entropy evolution for both models with $M=20$ for various initial conditions. Again $S_{\rm peak}\approx S_{\rm max}$ is found for highly excited initial states.

\begin{figure}[!htp]
\includegraphics[width=0.99\columnwidth]{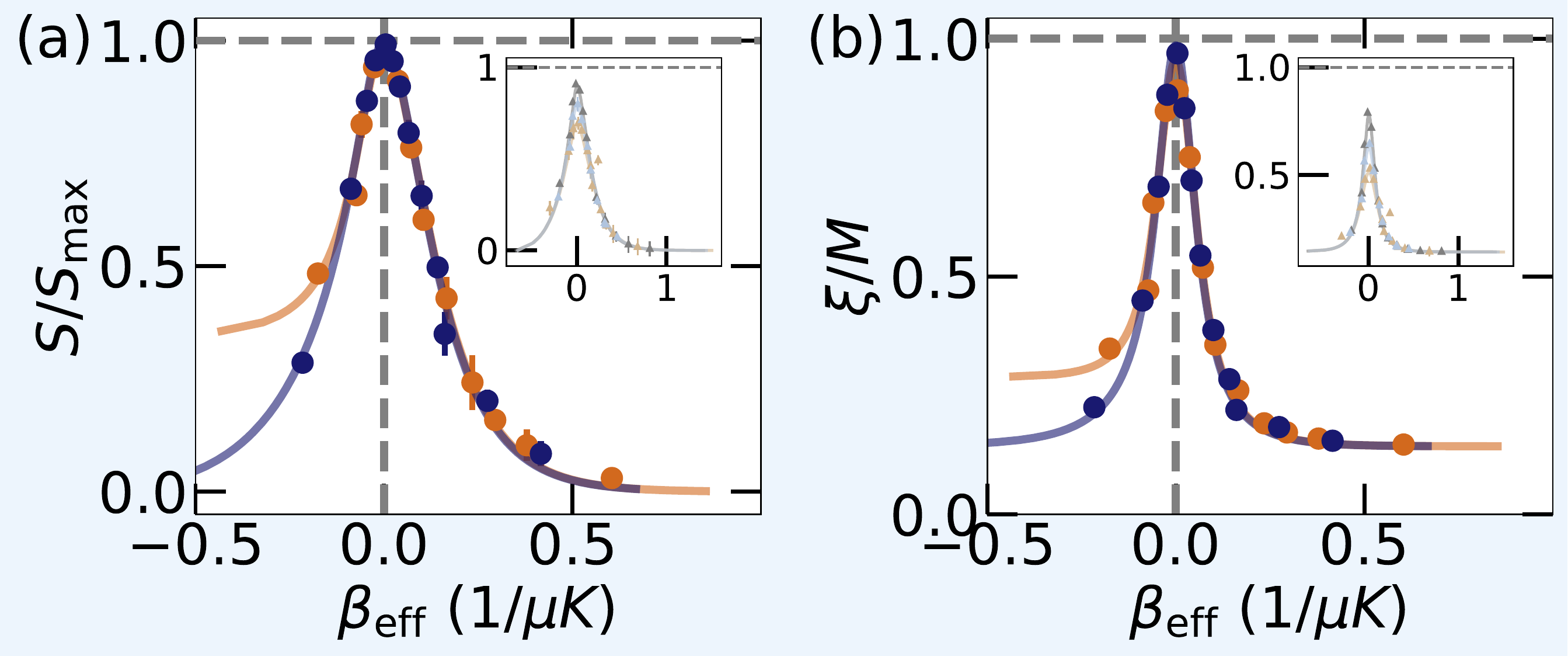} 
\includegraphics[width=0.99\columnwidth]{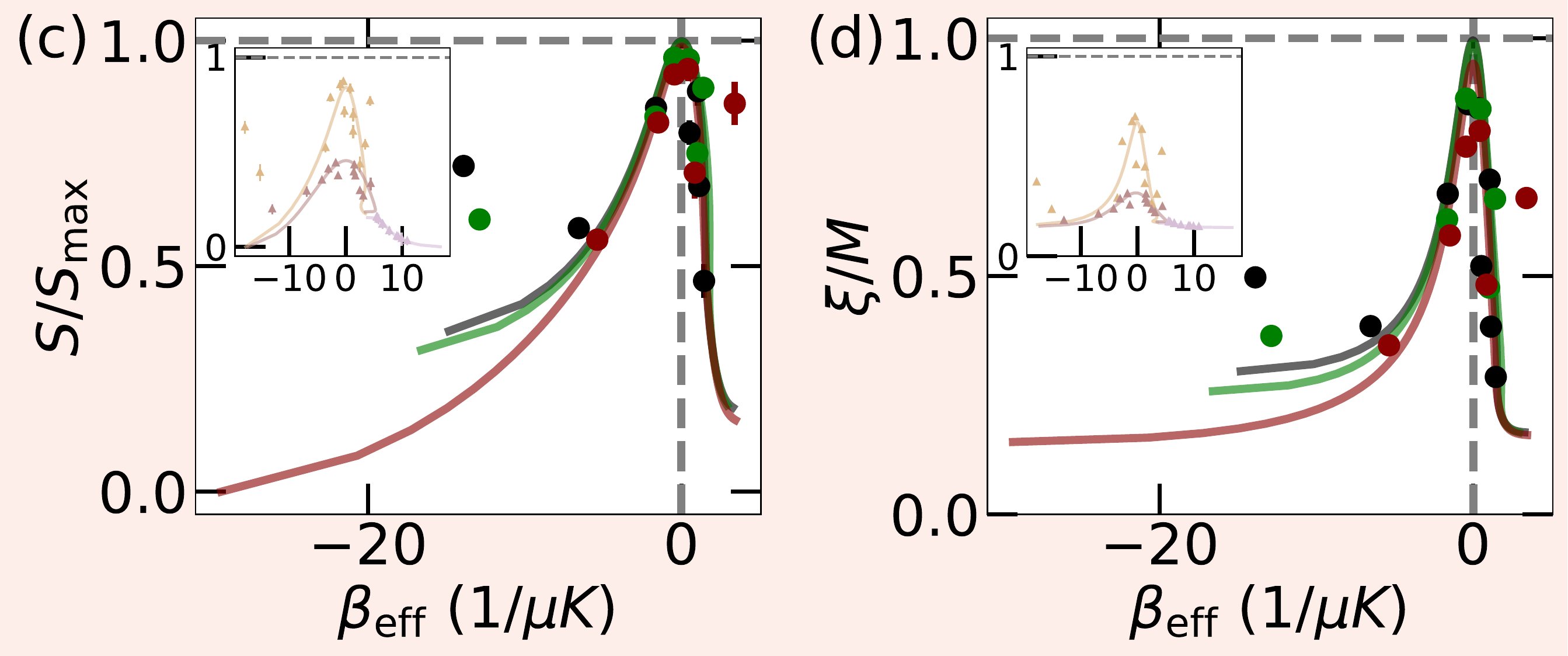}  
	\caption{{\bf Control parameter.}  Experimentally measured (symbols) and simulated (lines) entropy (a,c) and localization length (b,d) plotted as a function of $\beta_{\rm eff}$ [data, colors, and symbols like in Fig.~\ref{entropy_evolution}(a,b)]. \textcolor{black}{The main (inset) panel shows the results for the initial conditions far from (close to) equilibrium. The errorbars for $\xi/M$ are too small to be seen.} 
	}
	\label{beta_eff}
\end{figure}

To compare data for different initial conditions, we introduce the {scaled control parameter}
\begin{eqnarray}
	\beta_{\rm eff} \equiv \frac{dS/dt}{dE/dt}=\frac{dS}{dE},
\end{eqnarray} 
with mean energy $E$, {having the dimension of an inverse temperature}. It is monotonically related to the time $t$~(see supplemental material for more details) and becomes zero at $t=t_{\rm peak}$, while it takes negative (positive) values for $t<t_{\rm peak}$ ($t>t_{\rm peak}$). Despite superficially resembling an effective inverse temperature, we would like to stress that the use of this parameter does not imply that the system assumes a Gibbs-like state with effective time-dependent inverse temperature $\beta_\text{eff}$ during its transient evolution.
In Figs.~\ref{beta_eff}(a) and (c), the measured entropy is plotted as a function of $\beta_{\rm eff}$. For those initial conditions giving rise to close-to-maximum peak entropies, marked by bullets, the data collapse in the vicinity of $\beta_{\rm eff}=0$. Such behavior is equally visible, when plotting the scaled localization length $\xi/M$ [Figs.~\ref{beta_eff}(b) and (d)]. The insets show results for initial conditions corresponding to less excited states, for which the data does not collapse.

\begin{figure}[!htp]
    \includegraphics[width=0.99\columnwidth]{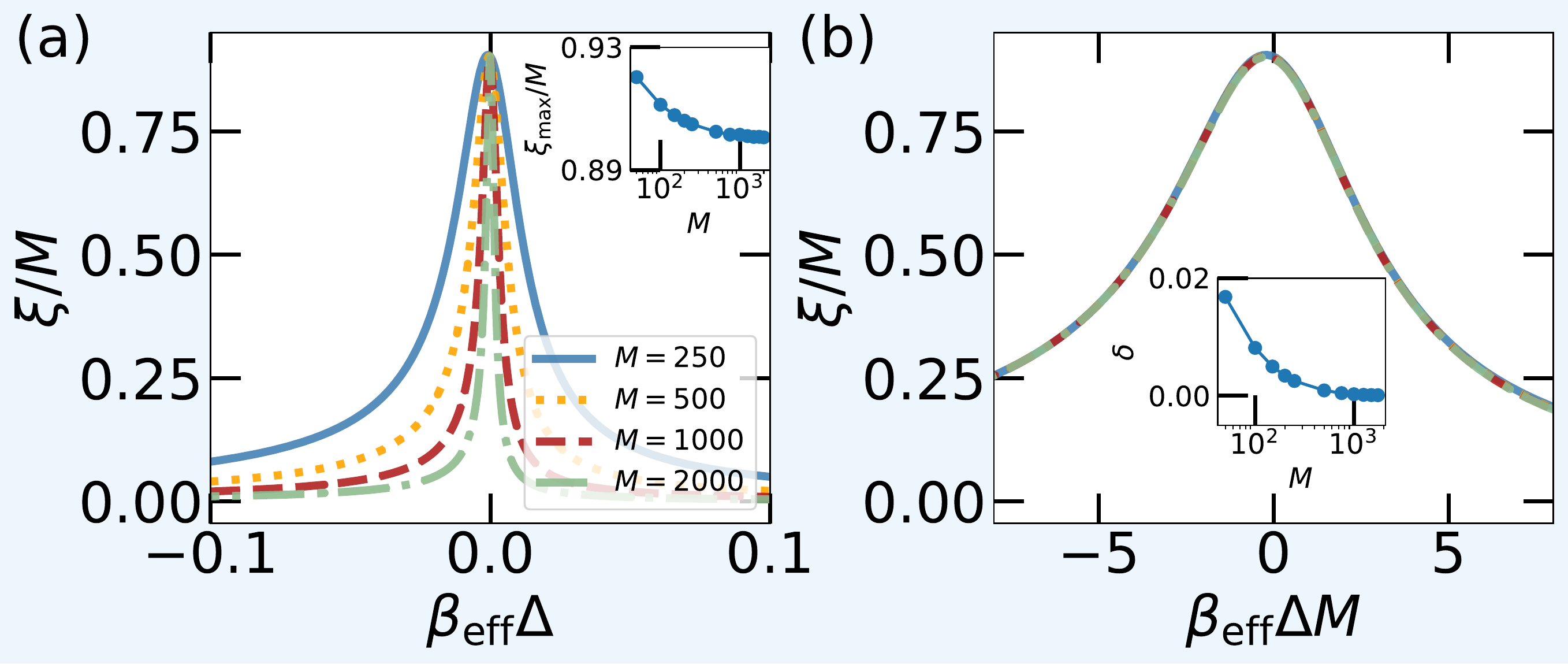}  \includegraphics[width=0.99\columnwidth]{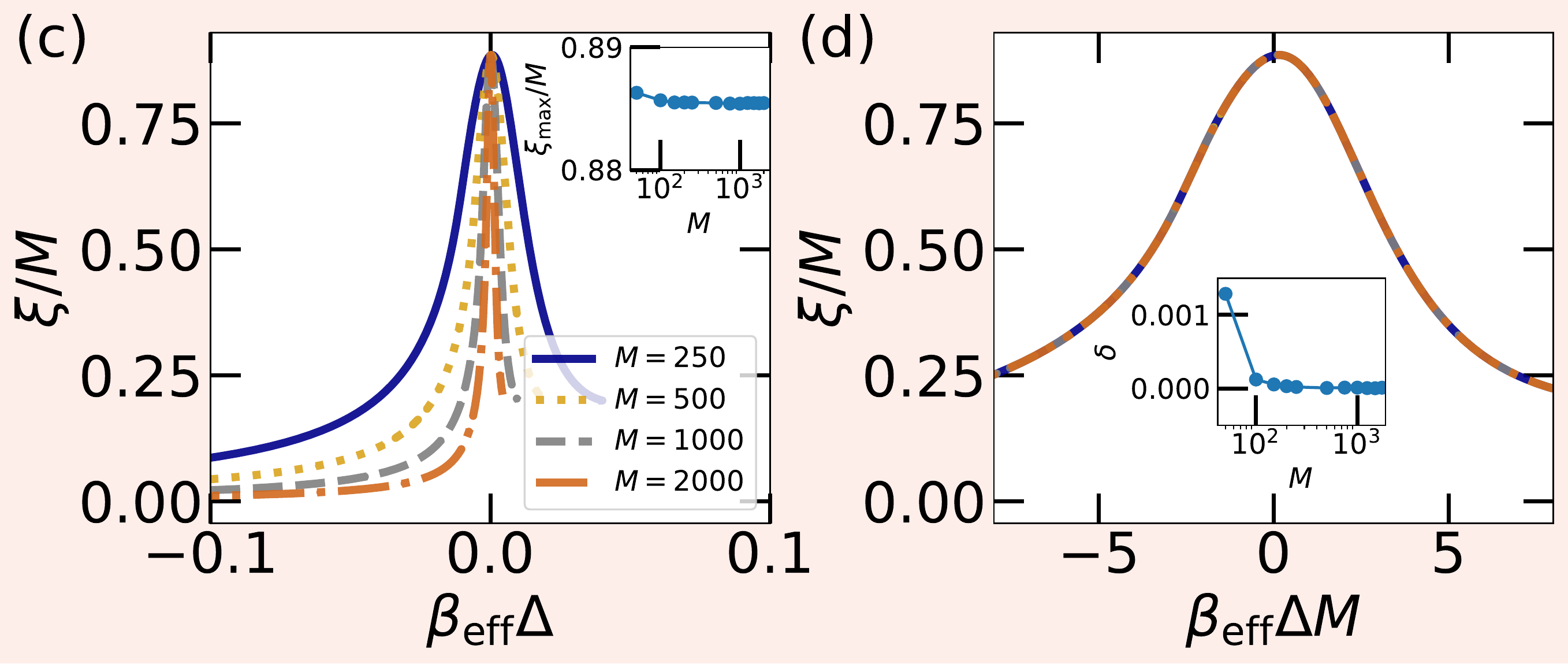}
	\caption{{\bf Finite-size scaling.} Scaled Localization length $\xi/M$ for different system sizes $M$ for the unidirectional (a,b) and the bidirectional (c,d) model as a function of $\beta_{\rm eff}\Delta$ (a,c) and the scaled parameter $\beta_{\rm eff}\Delta M$ corresponding to $\nu=1$ (b,d). The insets show the maximal $\xi$ as a function of the system size and the mean distance $\delta$ of the data for system size $M$ with that for system size $M=2000$.}
	\label{xi}
\end{figure}

{
Assuming a continuous phase transition in the thermodynamic limit, $M\to\infty$, the localization length in this limit, $\xi_\infty$, is expected to diverge like $\xi_\infty \propto \beta_{\rm eff}^{-\nu}$ at the transition point $\beta_\text{eff}=0$, where $\nu$ is a critical exponent.} As a consequence, the behavior of a large finite system should asymptotically depend on the ratio of $\xi_\infty$ and the system size~$M$ only, or, equivalently on $(\xi_\infty/M)^{-1/\nu}=\beta_{\rm eff} M^{1/\nu}$~\cite{domb2000phase}. In particular, the length $\xi$ for a system of finite large size $M$ is expected to behave as $\xi/M=g(\beta_{\rm eff} M^{1/\nu})$ close to the transition point $\beta_\text{eff}=0$, with some scaling function $g$. In Fig.~\ref{xi} we show $\xi/M$ for different system sizes $M$ as a function of both the dimensionless control parameter $\Delta\beta_\text{eff}$ as well as the scaled control parameter $\Delta\beta_\text{eff} M$, corresponding to a critical exponent of $\nu=1$. Remarkably, in the latter case we find an almost perfect collapse of the data, suggesting universal scaling as it is found at a continuous phase transition.  
We note that the peak of $\xi/M$ does not fully reach $1$. However, since the maximum of $\xi/M$ remains constant with increasing system size, $\xi$ diverges at $t_\text{peak}$ in the thermodynamic limit. These observations suggest the interpretation as a phase transition with respect to time. 

{The universal behaviour of the dynamics observed in the large-system limit can be understood better by mapping the thermodynamic limit to a continuum limit: For a hypothetical system of fixed size $L$, the variable $x=Lm/M$ becomes continuous for $M\to\infty$ and the rate equation for the probability distribution $p_m$ approaches a differential equation for the probability density $\rho(x)=(M/L)p_{xM/L}$. Close to the transition, the observed exponent $\nu=1$ can then be explained by starting from a maximally delocalized distribution, $\rho(x)=1/L$, and studying the evolution for small positive and negative times perturbatively. More details can be found in the supplemental material, where we also present further analyses, showing universal scaling of the specific-heat-like quantity $C=dE/d(\beta_\text{eff}^{-1})$, and discussing the phase transition with time $t$ (rather than $\beta_\text{eff}$) playing the role of the control parameter.}

    \begin{figure}[!htp]
    \includegraphics[width=1\columnwidth]{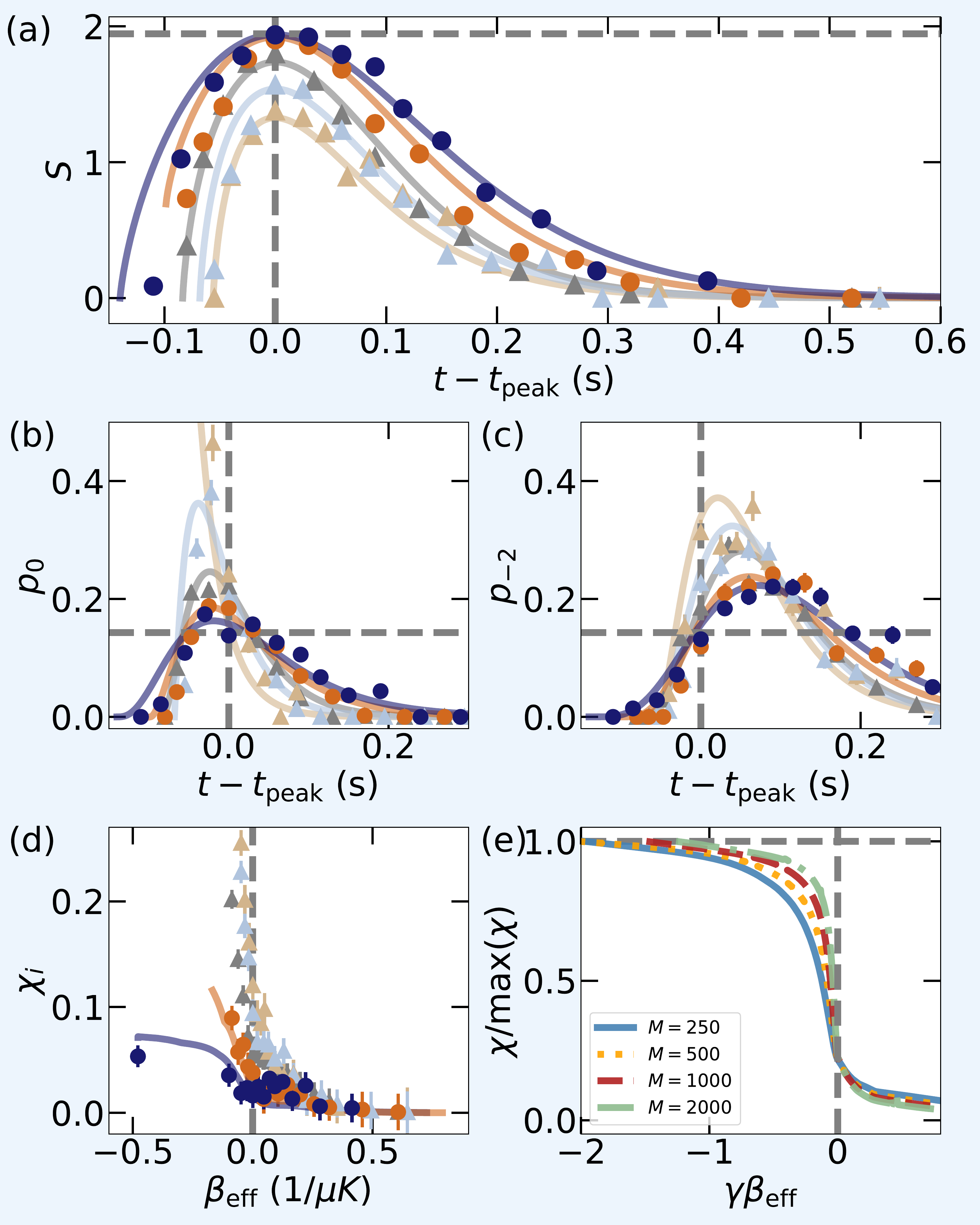}
	\caption{{\bf Prethermal memory loss.} (a) Experimentally measured~(symbols) and simulated (lines) entropy for different initial conditions (color coding as in Fig.~\ref{entropy_evolution}(a)) and (b, c) the corresponding population $p_{m_F}$ for the two spin states (b) $|m_F = 0\rangle$ and (c) $|m_F = -2\rangle$ as a function of shifted time (by the peak entropy time $t_{\rm {peak}}$) for the unidirectional model. The dependency of $\chi$~(see text for the definition) on the control parameter $\beta_{\rm eff}$ is shown in (d) for the experiments and in (e) for the theoretical model. 
	Horizontal grey dashed lines mark the maximal entropy $S_{\rm max}$ in (a), the population $1/7$ in (b, c) that corresponds to $S_{\rm max}$, and 1 in (e). Vertical dashed lines mark $t=t_{\rm {peak}}$ and $\beta_{\rm eff}=0$, respectively.
	}
	\label{entropy_odering_para}
    \end{figure}
    

Since the maximum entropy $S_{\rm max}$ corresponds to a unique state, the maximally mixed state $\rho_{\rm max}$, after approaching {$S\approx S_{\rm max}$}, the dynamics is expected to become {(approximately)} independent of the details of the initial state. This happens long before the spin system has thermalized~\cite{Wu2020}. Such a prethermal memory loss is observed in the experiment for both magnetic field regimes. In Figs.~\ref{entropy_odering_para}(a)-(c), we show the entropy evolution and the population dynamics of two spin states versus the shifted time $t-t_{\rm peak}$ in the regime of unidirectional rates (corresponding plots for the bidirectional regime are presented in the supplemental material).  We can see that the data with a high peak entropy ($S_{\rm peak}\geq 0.98 S_{\rm max}$, indicated by bullets) show similar behavior for both the entropy evolution and spin dynamics after the system reaches peak entropy~($\tau \equiv t-t_{\rm {peak}}>0$). 

To quantify this observation, we introduce the distance
\begin{equation}
	\chi_{ij}(\tau)=\frac{1}{M} \sum_{m=1}^M|p_{m}^{(i)}(t^{(i)}_\text{peak}+\tau) - p_{m}^{(j)}(t^{(j)}_\text{peak}+\tau)|
\label{eq_chi}
\end{equation}
between two trajectories with different initial conditions, $p^{(i)}_{m_F}(0)$ and $p^{(j)}_{m_F}(0)$, and peak times $t_\text{peak}^{(i)}$ and $t_\text{peak}^{(j)}$.
For the experimental data we compare each trajectory $p_{m_F}^{(i)}$ to the optimal trajectory $p_{m_F}^{\rm max}(t)$ defined by $S_{\rm peak}=S_{\rm max}$. 
In Fig.~\ref{entropy_odering_para}(d), we plot the corresponding distance $\chi_i\equiv \chi_{i\text{max}} $ versus $\beta_{\rm eff}$.
For those trajectories featuring large peak entropies, $\chi_i$ becomes small at the transition $\beta_{\rm eff}=0$. 
In comparison, for trajectories with $S_{\rm peak}<0.98 S_{\rm max}$ (indicated by triangles) $\chi_i$ remains large after the transition. 

Prethermal memory loss is also found in the theoretical models. Here we have easy access to many initial conditions and, therefore, we can compute the mean distance $\chi\equiv\mathrm{mean}_{ij\in\mathcal{U}}(\chi_{ij})$ of those trajectories whose peak entropies $S_{\rm peak}$ are close to the maximal entropy, i.e.,\ for which $S_{\rm peak}/S_{\rm max}>1-\delta S$ with threshold $\delta S\ll1$. 
Fig.~\ref{entropy_odering_para}(e) plots normalized $\chi$ versus $\beta_{\rm eff}$ for different system sizes with $\delta S=0.2$. One can see that for increasing $M$ a sharp transition forms at $\beta_{\rm eff}=0$.

In summary, we have investigated the far-from-equilibrium relaxation dynamics of an open quantum system given by a large spin coupled to a bath. 
We find that for highly excited initial states, the system transiently approaches the maximally mixed state $\rho_{\rm max}$, as signaled by a peak in the entropy evolution approximately reaching the maximally possible value $S_{\rm max}$. 
We show that, when reaching the entropy peak, the dynamics shows distinct features that signal critical scaling with respect to time: (i) In the limit of large system sizes, the localization length $\xi$ characterizing the spin state, diverges at the transition. (ii) A finite-size scaling analysis reveals a power-law scaling $\xi\sim \beta_{\rm eff}^{-\nu}$ near the transition, with respect to the scaled control parameter $\beta_{\rm eff}$, which is monotonically related to time and allows to compare data for different initial states by locating the transition to $\beta_{\rm eff}=0$. (iii) The extracted critical exponent takes the same value {$\nu=1$} for all model parameters considered, suggesting universal scaling behavior independent of the microscopic details of the system. Thus, we conclude that  critical behavior with respect to time can not only occur in the evolution of isolated systems described by pure states, but also during the dynamics of an open system induced by dissipation.  
It will be interesting to further investigate the nature of {such}  dynamical critical scaling in open quantum systems, including  its non-equilibrium universality classes (to the exploration of which our results provide a first step and a new approach). 
Another subject for future theoretical and experimental exploration is the collective behaviour of many atoms in contact with the bath as it results both from quantum statistics as well as from potential interactions. Also the regime of stronger system-bath coupling, where non-markovian effects are expected, offers an intriguing perspective. 

We thank Markus Heyl, Eric Lutz, and Li You, for their useful comments on the manuscript.
This work was supported by the Deutsche Forschungsgemeinschaft (DFG, German Research Foundation) via the Collaborative Research Centers SFB/TR185 (Project No. 277625399) and SFB 910 (Project No. 163436311). S.B.\ acknowledges funding from Studienstiftung des deutschen Volkes.

\bibliography{bibliography}{}

\clearpage


\section*{Methods}
\subsubsection*{Initial state preparation}\label{prep}
Experimentally, the Rb bath is prepared by laser-cooling in a magneto-optical trap (MOT) and subsequent cooling by evaporation while the sample is trapped in a crossed dipole trap at a wavelength of $\lambda=\SI{1064}{\nano m}$.
The bath's internal state is prepared via an optical pumping in $\ket{F_{\mathrm{Rb}}=1, m_{F,\mathrm{Rb}}=1}$ and then transferred via the radio-frequency transition $\ket{F_{\mathrm{Rb}}=1, m_{F,\mathrm{Rb}}=1} \rightarrow \ket{F_{\mathrm{Rb}}=1, m_{F,\mathrm{Rb}}=0}$ to a magnetic-field insensitive state. 
This allows us to accumulate Cs atoms from the atomic background vapor by laser cooling in a MOT only approximately $\SI{200}{\mu m}$ apart from the Rb sample. Subsequently, a crossed dipole trap with a wavelength of $\lambda$ loads the atoms from the MOT. Degenerate Raman sideband cooling~\cite{Kerman_2000} reduces the Cs temperature further while at the same time populating the bare atoms' absolute ground state $\ket{F_{\mathrm{Cs}}=3, m_{F,\mathrm{Cs}}=3}$.
Microwave-driven Landau-Zener transitions near-resonant to the $\ket{F_{\mathrm{Cs}}=3} \rightarrow \ket{F_{\mathrm{Cs}}=4}$ hyperfine transition ($h \times \SI{9.1}{GHz}$) prepare the Cs atoms in the desired initial state.

The interaction between Cs and Rb is initialized by transporting the Cs atoms into the bath via a species-selective optical lattice~\cite{Schmidt_2016}. The interaction stops after applying a resonant laser pulse that pushes the Rb atoms out of the trap. Eventually, state-selective fluorescence imaging~\cite{Schmidt_2018a} yields the internal state and position of the Cs atoms. 
\\
\subsubsection*{Experimental parameters}\label{exp_para}
The bath temperature $T$ and density $n$ for each measurement are inferred from time-of-flight measurements of the Rb cloud on the one hand; and from comparing the seven measured $m_F$-state trajectories with hundreds of simulated state trajectories on the other hand. Each simulation contains slightly different bath parameters. The bath parameters yielding the smallest least-squares ($\chi^2$) error for all trajectories and the independent time-of-flight measurement was used for the respective measurement data set.
The individual parameters of each measurement and the corresponding initial population are listed in the supplemental material.
For simplicity, Table \ref{tab:Experiment Parameter} shows the mean temperature and mean density of all best-fitting parameters for the unidirectional, respectively, bidirectional experimental system. 
Moreover, the magnetic field is calibrated via microwave spectroscopy on the $\ket{F_{\mathrm{Rb}}=1, m_{F,\mathrm{Rb}}=0} \rightarrow \ket{F_{\mathrm{Rb}}=2, m_{F,\mathrm{Rb}}=1}$ transition of the Rb bath.

\begin{table}[!htp]
\centering
\renewcommand{\arraystretch}{1.5} 
 \caption[Experimental Parameter]{Experimental parameters and ratio of mean rates}
\begin
{tabular}{lcc} 
Parameter & Unidirectional & Bidirectional  \\ 
\hline\hline

$B$ [mG] &  $\SI{460\pm2}{}$\  &  $\SI{25\pm2}{}$   \\ 
$T$ [nK]  & $\SI{920\pm24}{}$  & $\SI{492\pm31}{}$ \\ 
$n$ [$10^{13}\mathrm{cm}^{-3}$] &  $\SI{0.46\pm0.02}{}$  & $\SI{0.51\pm0.09}{}$  \\ 
$R_{+,m_F}/R_{-,m_F}$ & $\approx 10^{-5}$    & $0.21$   \\ 
\end{tabular}
\label{tab:Experiment Parameter} 
\end{table}
\subsubsection*{Inter-species spin-exchange processes}\label{exp_para}

The Zeeman energy for a bare Cs atom reaches its minimum for $\ket{m_F=3}$, defining the single-atom ground state. 
However, the situation reverses when the Cs atom is immersed in a bath of Rb atoms in the $\ket{m_{F,\mathrm{Rb}}=0}$ state. 
For this Rb-Cs combination, spin-exchange collisions can exchange one quantum of angular momentum between one atom of the bath and the Cs atom while the total angular momentum is preserved. 
At the same time, Zeeman energy is exchanged. Due to different atomic Landé factors, the Zeeman splitting of Rb is twice the splitting of Cs. Therefore, the spin- and energy exchange direction is essential and corresponds to two complementary processes in the bath. 
The process $\vert m_F^{\mathrm{Cs}}, m_F^{\mathrm{Rb}}\rangle \rightarrow \vert m_F^{\mathrm{Cs}}-1, m_F^{\mathrm{Rb}}+1\rangle$ is exoergic, and the energy amount corresponding to one Cs atom's Zeeman energy $\hbar \Delta = \mu_B g_F^\mathrm{Cs} B$ is released as kinetic energy and dissipated by subsequent elastic collisions in the bath. 
The complementary process $\vert m_F^{\mathrm{Cs}}, m_F^{\mathrm{Rb}}\rangle \rightarrow \vert m_F^{\mathrm{Cs}}+1, m_F^{\mathrm{Rb}}-1\rangle$ is endoergic, and the kinetic collisional energy of the Cs atom and bath atom must provide the energy amount $\hbar \Delta$ for this collision to occur.
The collisional energy is Maxwell-Boltzmann distributed. For the ultracold temperatures of the bath, the rates for exothermal and endothermal SE collisions, $R_-$ and $R_+$, respectively, have markedly different rates with $R_- > R_+$.
As a consequence, the definitions of ground and highest excited states invert, and the former bare-atom ground (highest-excited) state, i.e., $\ket{m_F=+3}$ ($\ket{m_F=-3}$), defines the impurity's highest excited (ground) state.

\subsubsection*{Spin evolution calculation}
The evolution of the probability in eigenstate $|m\rangle$, $p_m$, is described by the rate equation
\begin{eqnarray}
	\dot p_m = R_{+,m-1} p_{m-1} + R_{-,m+1} p_{m+1} - (R_{-,m} + R_{+,m}) p_m. \notag
\end{eqnarray}
where $R_{\pm, m} \equiv R_{m\pm1,m}$ denotes the transfer rate from  eigenstate $|m\rangle$ to eigenstate $|m\pm1\rangle$. For the unidirectional model discussed in the main text, $R_{+,m}=0$. For the bidirectional model with state-independent rates, $R_{\pm,m} \equiv R_\pm$.

For simulating the experimental spin dynamic, the rates are given by $R_i = \langle n \rangle \sigma_i(B,T) \bar{v}$, with $i= m_F\pm1,m_F$, mean relative velocity of the colliding atoms $\bar{v}$, Cs-Rb density overlap $\langle n \rangle$ and state-dependent scattering crossing section $\sigma_i$. The ratio of the mean rates $R_{+,m_F}/R_{-,m_F}$ in Table~\ref{tab:Experiment Parameter} shows an experimentally accurately blocking of the endothermal rates $R_{+,m_F}$ by choice of a large magnetic field. 


\section*{Data availability}~\\
All data supporting the finding of this paper are available from the corresponding author A.W. upon reasonable request.

\section*{Code availability}~\\
The codes that support the findings of this paper are available from the corresponding author A.E. upon reasonable request.

\clearpage

\noindent {\LARGE \bf Supplementary Materials:}\\

\setcounter{equation}{0}
\setcounter{figure}{0}
\setcounter{table}{0}
\setcounter{page}{1}
\renewcommand{\theequation}{S\arabic{equation}}
\renewcommand{\thefigure}{S\arabic{figure}}
\renewcommand{\thetable}{S\arabic{table}}
\renewcommand{\bibnumfmt}[1]{[S#1]}

\noindent {\bf Prethermal memory loss.}\\


    \begin{figure}[!htp]
	    \includegraphics[width=1\columnwidth]{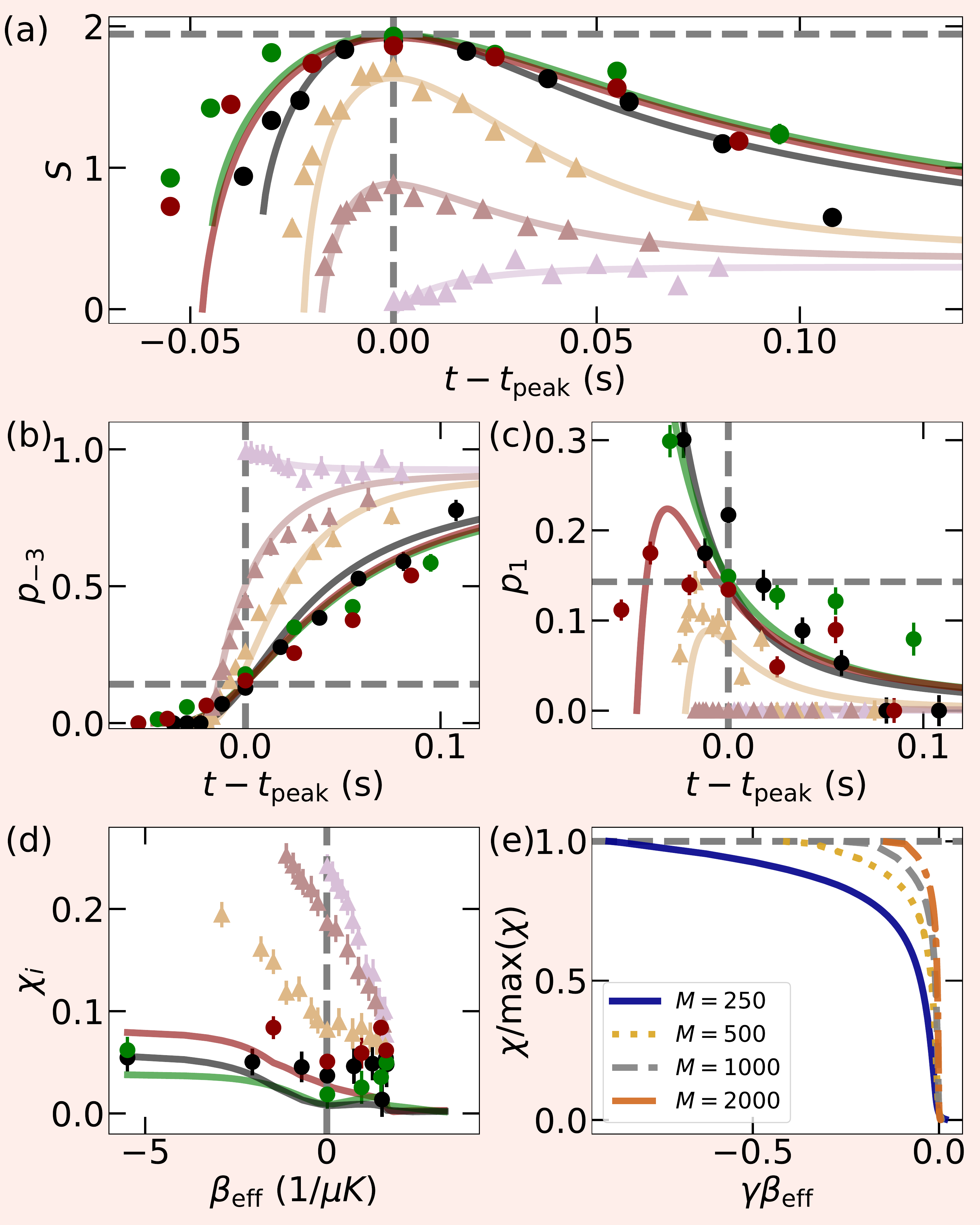}
		\caption{{\bf Prethermal memory loss in the bidirectional model with state-independent rates.} (a) Experimentally measured~(symbols) and simulated (lines) entropy for different initial conditions~(color coding as in Fig.~\ref{entropy_evolution}(b) of the main text) and (b, c) the corresponding spin population of state $|m_F\rangle$, $p_{m_F}$, as a function of shifted time (by the peak entropy time $t_{\rm {peak}}$) for the bidirectional model. As an illustration, we show two spin components. The dependency of $\chi$~(see the main text for the definition) on the control parameter $\beta_{\rm eff}$ is shown in (d) for the experiments and in (e) for the theoretical model. 
			Horizontal grey dashed lines mark the maximal entropy $S_{\rm max}$ in (a), the population $1/7$ in (b, c) that corresponds to $S_{\rm max}$, and 1 in (e). Vertical dashed lines mark $\beta_{\rm eff}=0$.
			}
		\label{DD_odering_para}
	    \end{figure}


In the main text, the prethermal memory loss was illustrated for the unidirectional model in Fig.~\ref{entropy_odering_para}. 
Figure~\ref{DD_odering_para} shows the complementary data for the bidirectional model with state-independent rates used in the theoretical model. 
In Figs.~\ref{DD_odering_para}(a)-(c), we show the entropy evolution and the population dynamics of two spin states with respect to the shifted time $t-t_{\rm peak}$. We can see that the data with a high peak entropy ($S_{\rm peak}\geq 0.98 S_{\rm max}$, indicated by bullets) show similar behavior for both the entropy evolution and spin dynamics after the system reaches peak entropy~($t-t_{\rm {peak}}>0$). 
In Fig.~\ref{DD_odering_para}(d), the difference between trajectories $\chi_i$~(see the definition in the main text) is plotted versus control parameter $\beta_{\rm eff}$.
For those trajectories featuring large peak entropies~(indicated by bullets), $\chi_i$ becomes small at the transition $\beta_{\rm eff}=0$ (though the measured signal is not as clear as for the unidirectional model discussed in the main text), whereas it remains large at $\beta_{\rm eff}=0$ for the trajectories with $S_{\rm peak}<0.98 S_{\rm max}$ (indicated by triangles).
In Fig.~\ref{DD_odering_para}(e), we plot normalized $\chi$ versus $\beta_{\rm eff}$ for the theoretical bidirectional model with different system sizes $M$ at $\delta S=0.02$. One can clearly see that for increasing $M$, a sharp transition forms at $\beta_{\rm eff}=0$. 

\begin{figure*}
    \includegraphics[width=1\columnwidth]{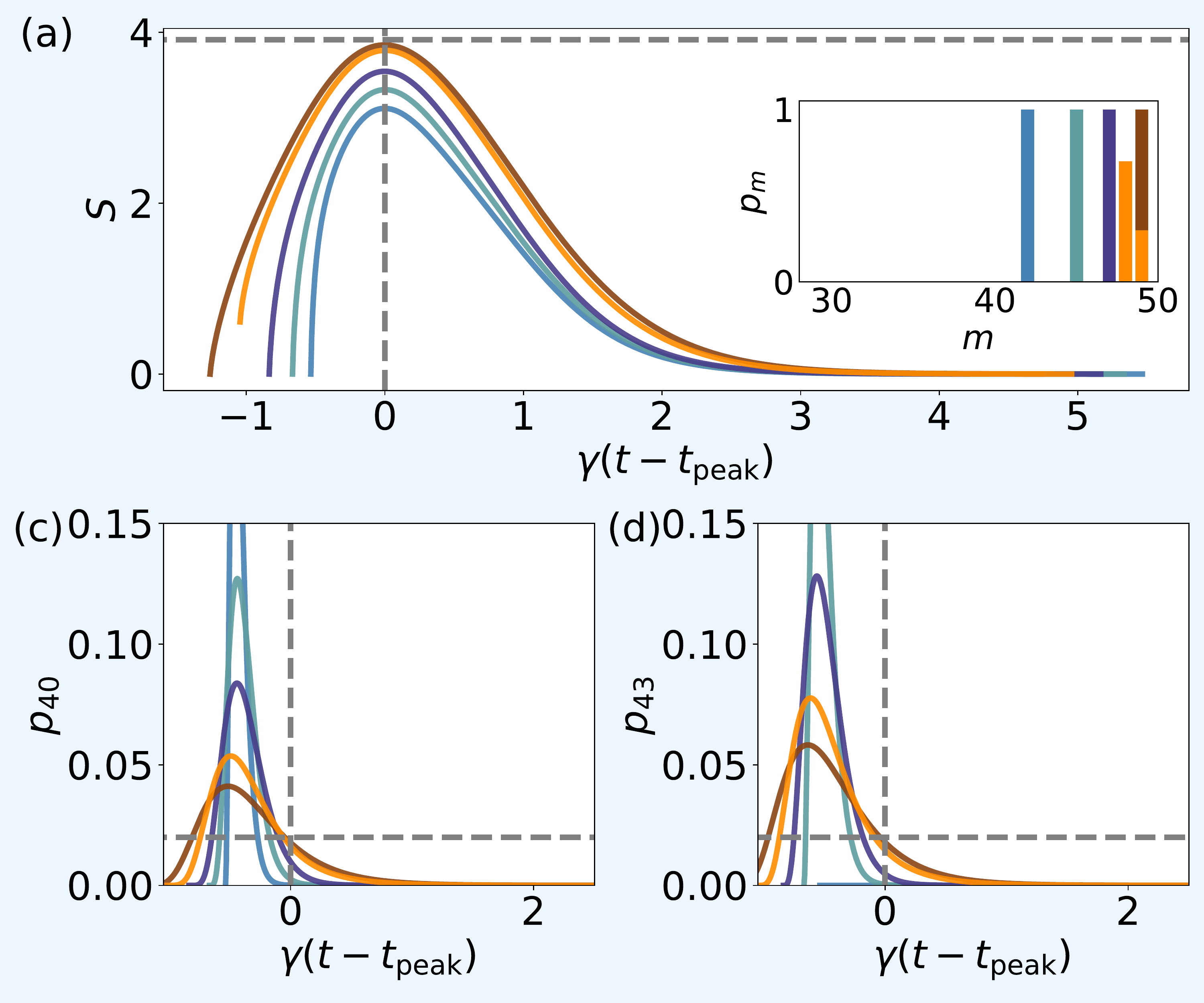}
    \includegraphics[width=1\columnwidth]{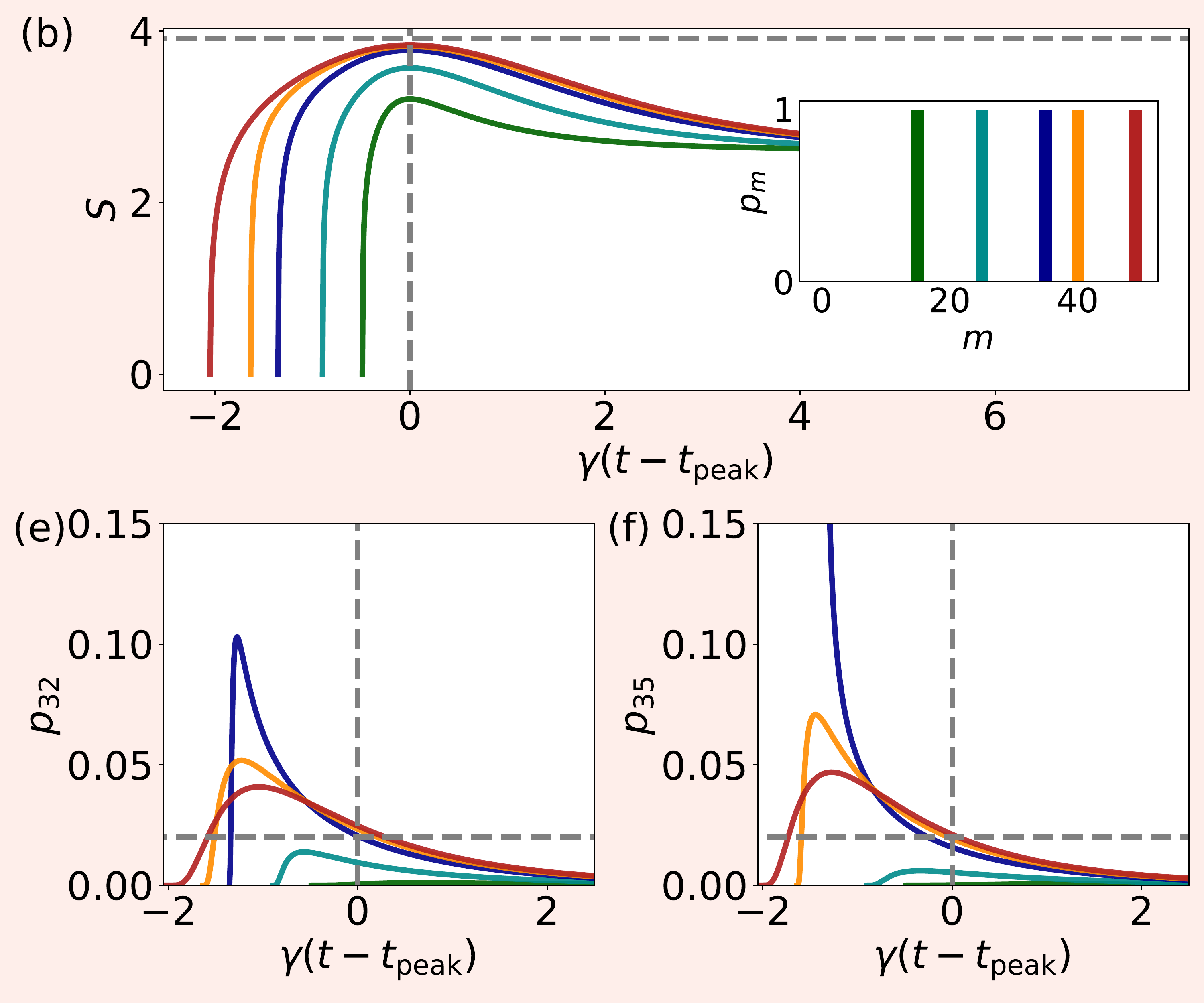}
\caption{{\bf Prethermal memory loss in the theoretical models.} (a,~b) Entropy $S$ and (c-f) populations on state $|m\rangle$, $p_{m}$, plotted as a function of shifted time~(by the peak entropy time $t_{\rm peak}$) for different initial conditions as shown in the inset. The horizontal dashed lines in (a,~b) mark the maximal entropy $S_{\rm max}=\ln M$ with $M=50$, in (c-f) mark $1/M=1/50$. The vertical dashed lines mark the peak entropy time $t=t_{\rm peak}$. Background colors are the same as figures of the main text: blue for the unidirectional model, red for the bidirectional model with state-independent rates.
		}\label{S_shift}
\end{figure*}

Prethermal memory loss is also found for both theoretical models, as shown in Fig~\ref{S_shift} for systems of size $M=50$. Panels (a) and (b) show the evolution of the entropy with respect to $t-t_\text{peak}$ for various initial conditions (as indicated by the insets). Panels (c-f) show the corresponding evolution for the populations of two different states for each model. One can clearly observe that (only) for those initial conditions for which the peak entropy closely approaches the maximum possible entropy [dashed line in panels (a) and (b)], the results for different initial conditions converge, when approaching $t=t_\text{peak}$, to remain very similar at all later times.


\begin{figure}[!h]
\centering
\includegraphics[width=0.99\columnwidth]{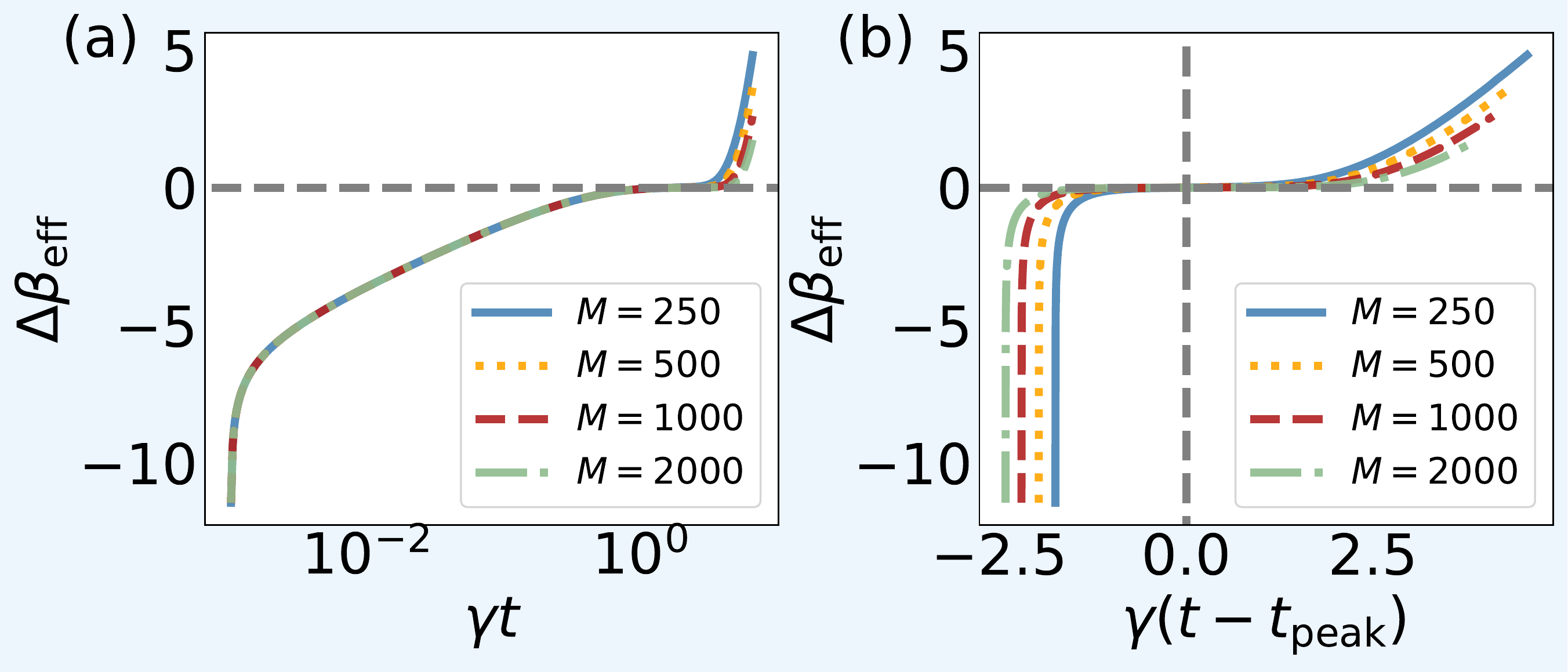}
\includegraphics[width=0.99\columnwidth]{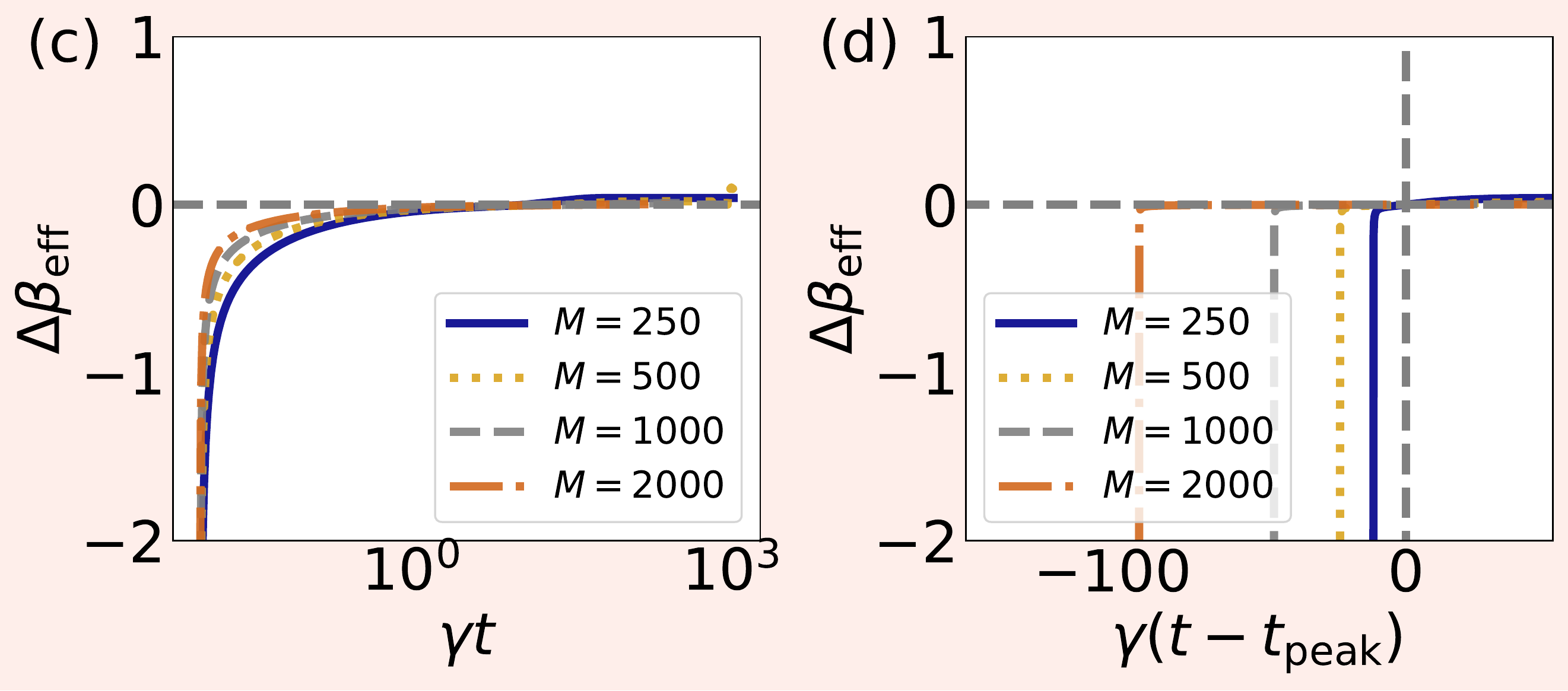}
	\caption{The control parameter $\beta_{\rm eff}$ as a function of time~(a,c) or shifted time~(b,d) for the unidirectional model~(blue background) and the bidirectional model~(red background).}
\label{beta_t}
\end{figure}

\noindent {\bf The mapping between the {control parameter} $\beta_{\rm eff}$ and time.}\\
Figure~\ref{beta_t} shows the control parameter $\beta_{\rm eff}$ as a function of time~(a,c) and shifted time~(b,d) for the unidirectional model~(blue background) and the bidirectional model.

\begin{figure}[!h]
\centering
\includegraphics[width=0.99\columnwidth]{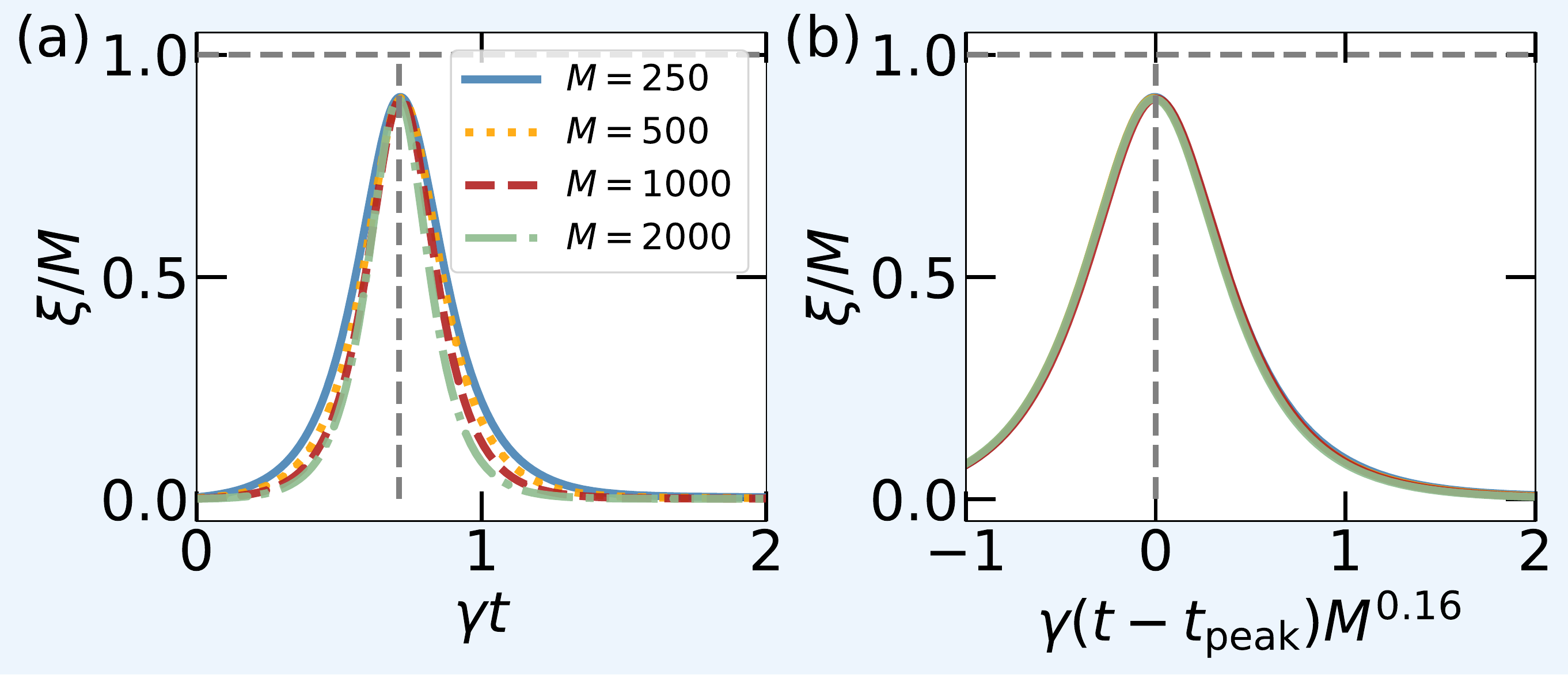}
	\caption{{\bf Finite-size scaling for the unidirectional model with time as control parameter.}
		(a) The normalized {localization length $\xi$} as a function of time for different system sizes. The overall strength of the bath-induced rates scales with the system size as $M^{1.16}$. The vertical dashed line marks the peak time $t_{\rm peak}$. (b) The normalized localization length $\xi$ as a function of the scaled shifted time (by $t_{\rm peak}$).}
\label{tFSS}
\end{figure}

\noindent {\bf Finite-size scaling for the unidirectional model with time as control parameter.}\\
{It is also possible to directly consider the time $t$, rather the {control parameter} $\beta_\text{eff}$ as the control parameter. In this case, one has to scale the overall strength of the bath-induced rates with the system size $M$ in such a way that the peak entropy time $t_\text{peak}$ remains finite in the thermodynamic limit, i.e.\ that it neither approaches infinity nor zero. For the  unidirectional model, we achieve this by scaling the rates like $M^{1.16}$. Note that such overall scaling of the rates, only changes the time scale of the evolution and not the details of the dynamics. (As a consequence, the dynamics plotted with respect to $\beta_\text{eff}$ remains unchanged, when choosing a different scaling). Under these conditions the divergence of both $\xi$ and $S$ occurs at (and within) the finite time $t_\text{peak}$ (see Fig.~\ref{tFSS}(a)) and, thus, constitutes singular behavior in time like at a continuous phase transition {in time}. In Fig.~\ref{tFSS}(b) we present a finite-size scaling analysis of this transition and find that the corresponding critical exponent is given by $1/0.16 = 6.25$. This is consistent with that found in the continuum model (see the discussion in section `Continuum model').}

\noindent {\bf Model of the experimental realization.}\\
\textcolor{black}{The experimental system comprises individual Cs atoms immersed in a large Rb bath. 
	The Hamiltonian of this Cs-Rb mixture is given by~\cite{Schmidt_Tailored}
	\begin{equation}
	    H = E_{\rm coll} + \sum\limits_{j={\rm Cs, Rb}}{(V_j^Z+V_j^{\rm HFS})+H_{\rm int}}.
	\end{equation}
	Here, $E_{\rm coll}$ denotes kinetic collision energy, $V_j^Z$ and $V_j^{\rm HFS}$ the single-particle Zeeman and hyperfine energies, respectively, and $H_{\rm int}$ describes the interaction of the colliding Rb and Cs atom. 
	The collision energy is well-defined by the relative velocity of the colliding partners for individual collisions, but in the ensemble it is distributed according to a Maxwell-Boltzmann distribution \cite{PhysRevD.89.103533}.  
	Considering low collision energies for the experimental ultra-low temperatures , i.e., the s-wave limit, the interaction $H_{\rm int}$ may be efficiently represented 
	in terms of asymptotic Cs (Rb) states, provided by total angular momentum $\bf{F}_{\rm Cs}$~($\bf{F}_{\rm Rb}$), with quantum numbers $F_{\rm Cs}$~($F_{\rm Rb}$) and projection $m_{F,\rm{Cs}}$~($m_{F,\rm{Rb}}$) resulting in \cite{Ho98}
	\begin{equation}
	H_{\rm {int}} = \sum\limits_{i=0,1,2} {c_i({\bf{F}}_{\rm {Cs}}\cdot {\bf{F}}_{\rm{ Rb}})^i}.
	\end{equation}
	In our experiments, Cs~(Rb) atoms are in hyperfine ground states $F_{\rm Cs}=3$~($F_{\rm Rb}=1$). 
	The interaction energy $H_{\rm int}$ becomes comparable to the hyperfine splitting $V^{\rm HFS}_j$ at distances of few $10 a_0$, with $a_0$ representing the Bohr radius, coupling $\bf{F}_{\rm Cs}$ and $\bf{F}_{\rm Rb}$. 
	The short spatial distance for the interaction justifies an effective contact interaction of the collision.
	This coupling can lead to different collision channels between Rb and Cs according to 
	\begin{equation}
	|m_{F,{\rm Cs}}',m_{F,{\rm Rb}}'\rangle=|m_{F,{\rm Cs}}+\Delta m_F,m_{F,{\rm Rb}}-\Delta m_F\rangle
	\end{equation}
	with $\Delta m_F=0, \pm 1, \pm 2$. 
	The system collides elastically for $\Delta m_F=0$, whereas the colliding atoms exchange angular momentum for $\Delta m_F \ne 0$.
	The latter processes comprise exoergic SE collisions where $\Delta m_F=-1,-2$ and endoergic SE collisions with $\Delta m_F=+1,+2$. 
	Due to the competition between Zeeman and thermal energy (for more details see \cite{Bouton20}), endoergic collisions are energetically forbidden for high magnetic fields, leading to a unidirectional spin-exchanging system. }

\noindent {\bf Markovianity of the experimental bath.}\\
In the experimental situation, Rb atoms are prepared in $m_{F,{\rm Rb}}=0$. 
In this case, angular momentum changes of $\Delta m_F=0, \pm 1$ are allowed. 
The number of Rb atoms is approximately three orders of magnitude larger than that of Cs atoms (see Tables S1 and S2). 
This strong imbalance between the bath~(Rb) and the probe~(Cs) together with the physical consequences of the Rb-Cs interaction justifies the assumption of an ideal Markov bath; i.e., Cs atoms exclusively interact with Rb atoms in the internal state~($m_{F,{\rm Rb}}=0$), and correlations by a second collision with the same Rb atom are negligible.
First, the rates for elastic collisions are more than a factor 30 larger than SE collisions (see Tables S1 and S2). Additionally, the Cs impurity after an inelastic collision needs approximately three elastic collisions to re-thermalize. Hence, for every SE collision the Cs impurity has the same bath temperature. 
Furthermore, the total number of atoms produced in state $m_{F,\mathrm{Rb}}\neq 0$ is very small during the interaction time (up to six Rb atoms per Cs impurity). Moreover, the Cs impurity undergoes more than 30 elastic collisions with Rb bath atoms and, furthermore, the mean-free path for our Cs atoms (averaged over all internal states) is larger than $9\,\mu$m ($7\,\mu$m) for the unidirectional (bidirectional) case, while the Rb bath has extensions of $31\,\mu$m ($23\,\mu$m) in axial and $3.2\,\mu$m ($2.4\,\mu$m) in radial direction, respectively. Thus, the Cs impurity samples the whole Rb cloud several times before another SE collision occurs. 
The probability of colliding with a Rb atom which had previously collided once with the Cs atom is therefore well below the percent level.
Hence, experimentally each SE collision occurs with identical conditions involving a Rb atom from an unchanged bath as originally prepared; this scenario realizes the ideal limit of a Markov bath and allows us modelling the time evolution of Cs atoms (driven by SE with Rb) based on a rate equation~(see Methods).

\noindent {\bf Experimental parameters.}\\
Tables S1 and S2 provide the experimental parameters for each data set of the unidirectional and bidirectional model realized at magnetic fields of $460\,$mG and $25\,$mG, respectively. The initial population of each data set is illustrated and color-coded as histogram in Fig.~\ref{entropy_evolution} of the main text. For simplicity the tables contain the two most populated initial states. The last missing percents of the initial population distribute over the unspecified states in the tables. Table~\ref{tab:Experiment Parameter} of Methods gives the mean values of the listed temperatures and densities.

\begin{table*}
	\small
	\centering
	\renewcommand{\arraystretch}{1.5} 
	\caption[Experimental Parameter individual measurements unidirectional]{Experimental parameters of the individual measurements for the unidirectional system, i.e. $B=\SI{460\pm2}{\milli\gauss{}}$.}
	\begin{tabular}{lcccccc}
		$p_{m_F}(t=0)$ & $T$ [nK] & $n$ [$10^{13}\mathrm{cm}^{-3}$] & $\frac{\Gamma_\text{SE}}{\Gamma_\text{elastic}}$ & $\frac{N_\text{Rb}}{N_\text{Cs}}$ & $\Gamma_\text{SE}$ [Hz] \\
		\hline\hline
		$p_{0}=\num{1.00\pm0.04}$ & \SI{900\pm75}{} & \SI{0.44\pm0.09}{} & 0.022 & 2300 & 8.969 \\
		\hline
		$p_{3}=\num{0.98\pm0.03}$ & \multirow{2}{*}{\SI{950\pm67}{}} & \multirow{2}{*}{\SI{0.44\pm0.08}{}} & \multirow{2}{*}{0.022} & \multirow{2}{*}{2595} & \multirow{2}{*}{8.82} \\
		$p_{2}=\num{0.02\pm0.01}$ & \\
		\hline
		$p_{2}=\num{0.90\pm0.02}$ & \multirow{2}{*}{\SI{950\pm52}{}} & \multirow{2}{*}{\SI{0.49\pm0.08}{}} & \multirow{2}{*}{0.022} & \multirow{2}{*}{2327} & \multirow{2}{*}{9.951} \\
		$p_{1}=\num{0.06\pm0.01}$ & \\
		\hline
		$p_{1}=\num{0.95\pm0.02}$ & \multirow{2}{*}{\SI{900\pm57}{}} & \multirow{2}{*}{\SI{0.45\pm0.06}{}} & \multirow{2}{*}{0.022} & \multirow{2}{*}{2411} & \multirow{2}{*}{9.218} \\
		$p_{0}=\num{0.05\pm0.01}$ & \\
		\hline
		$p_{2}=\num{0.51\pm0.03}$ & \multirow{2}{*}{\SI{900\pm52}{}} & \multirow{2}{*}{\SI{0.47\pm0.07}{}} & \multirow{2}{*}{0.022} & \multirow{2}{*}{2708} & \multirow{2}{*}{9.716} \\
		$p_{3}=\num{0.48\pm0.02}$ & \\
		\hline
	\end{tabular}
	\label{tab:Experiment_Parameter_sup1}
\end{table*}

\begin{table*}
	\small
	\centering
	\renewcommand{\arraystretch}{1.5}
	\caption[Experimental Parameter individual measurements bidirectional]{Experimental parameters of the individual measurements for the bidirectional system, i.e. $B=\SI{25\pm2}{\milli\gauss{}}$.}
	\begin{tabular}{lccccc}
		$p_{m_F}(0)$ & $T$ [nK] & $n$ [$10^{13}\mathrm{cm}^{-3}$] & $\frac{\Gamma_\text{SE}}{\Gamma_\text{elastic}}$ & $\frac{N_\text{Rb}}{N_\text{Cs}}$ & $\Gamma_\text{SE}$ [Hz] \\
		\hline\hline
		$p_{1}=\num{0.66\pm0.03}$ & \multirow{2}{*}{\SI{500\pm87}{}} & \multirow{2}{*}{\SI{0.41\pm0.16}{}} & \multirow{2}{*}{0.029} & \multirow{2}{*}{7086} & \multirow{2}{*}{9.581} \\
		$p_{2}=\num{0.25\pm0.02}$ & \\
		\hline
		$p_{2}=\num{0.78\pm0.03}$ & \multirow{2}{*}{\SI{450\pm43}{}} & \multirow{2}{*}{\SI{0.55\pm0.12}{}} & \multirow{2}{*}{0.030} & \multirow{2}{*}{6375} & \multirow{2}{*}{13.037} \\
		$p_{1}=\num{0.11\pm0.01}$ & \\
		\hline
		$p_{2}=\num{0.49\pm0.03}$ & \multirow{2}{*}{\SI{525\pm55}{}} & \multirow{2}{*}{\SI{0.63\pm0.12}{}} & \multirow{2}{*}{0.028} & \multirow{2}{*}{4591} & \multirow{2}{*}{14.509} \\
		$p_{1}=\num{0.44\pm0.02}$ & \\
		\hline
		$p_{0}=\num{0.85\pm0.03}$ & \multirow{2}{*}{\SI{500\pm47}{}} & \multirow{2}{*}{\SI{0.53\pm0.12}{}} & \multirow{2}{*}{0.029} & \multirow{2}{*}{4844} & \multirow{2}{*}{12.318} \\
		$p_{-1}=\num{0.07\pm0.01}$ & \\
		\hline
		$p_{-2}=\num{0.93\pm0.04}$ & \multirow{2}{*}{\SI{525\pm55}{}} & \multirow{2}{*}{\SI{0.38\pm0.11}{}} & \multirow{2}{*}{0.028} & \multirow{2}{*}{4086} & \multirow{2}{*}{8.831} \\
		$p_{-3}=\num{0.06\pm0.01}$ & \\
		\hline
		$p_{-3}=\num{0.99\pm0.04}$ & \multirow{2}{*}{\SI{450\pm54}{}} & \multirow{2}{*}{\SI{0.58\pm0.15}{}} & \multirow{2}{*}{0.030} & \multirow{2}{*}{3600} & \multirow{2}{*}{13.852} \\
		$p_{-2}=\num{0.01\pm0.01}$ & \\
		\hline
	\end{tabular}
\label{tab:Experiment_Parameter_sup2}	
\end{table*}

\begin{figure}[!h]
\centering
\includegraphics[width=0.9\columnwidth]{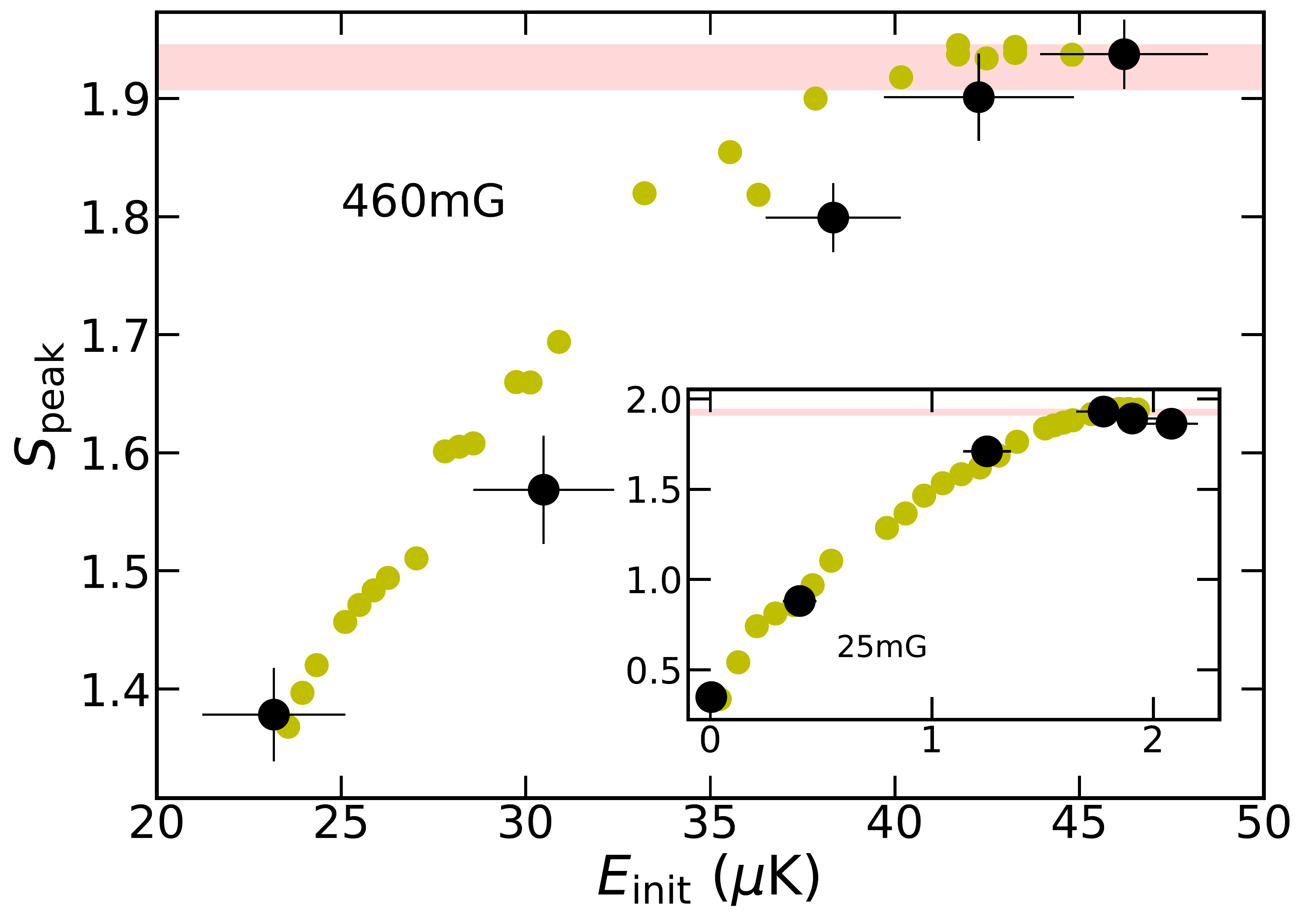}
	\caption{{\bf Peak entropy versus initial state energy.}
		The unidirectional and the bidirectional (inset) systems reach both maximum entropy ($S_{\rm peak}\ge 0.98 S_{\rm max}$ indicated by the red area) for a broad range of initial energies down to $\approx 85\%$ of the maximum possible energy before the peak entropy decreases. Big black markers illustrate data and small yellow dots simulations. Simulations consider the same temperature, magnetic field, and atom number, assuming some initial population distributions to sample
	    the initial energy.}
\label{SvsEinit}
\end{figure}

\begin{figure}[!h]
\centering
\includegraphics[width=0.49\columnwidth]{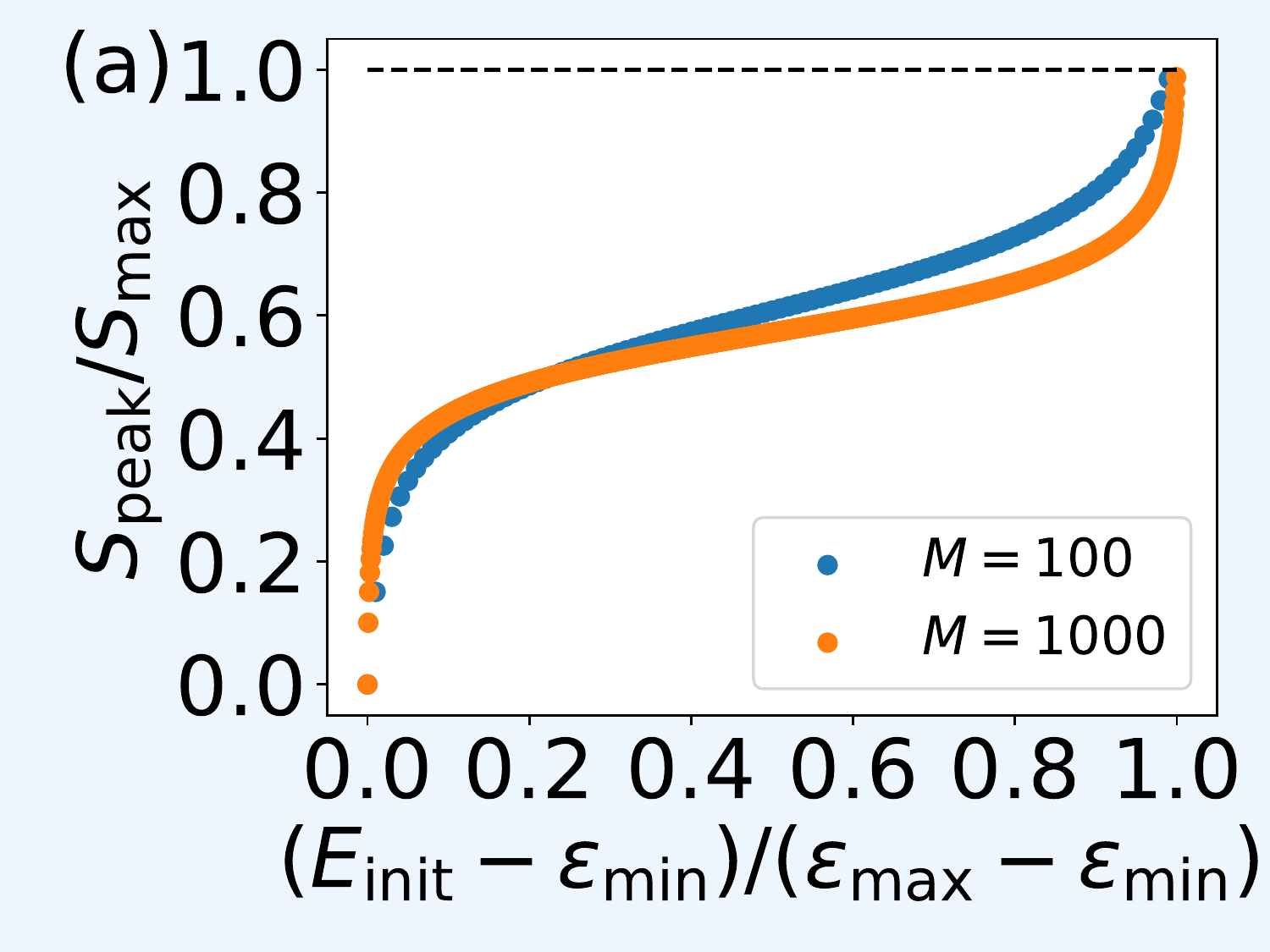}
\includegraphics[width=0.49\columnwidth]{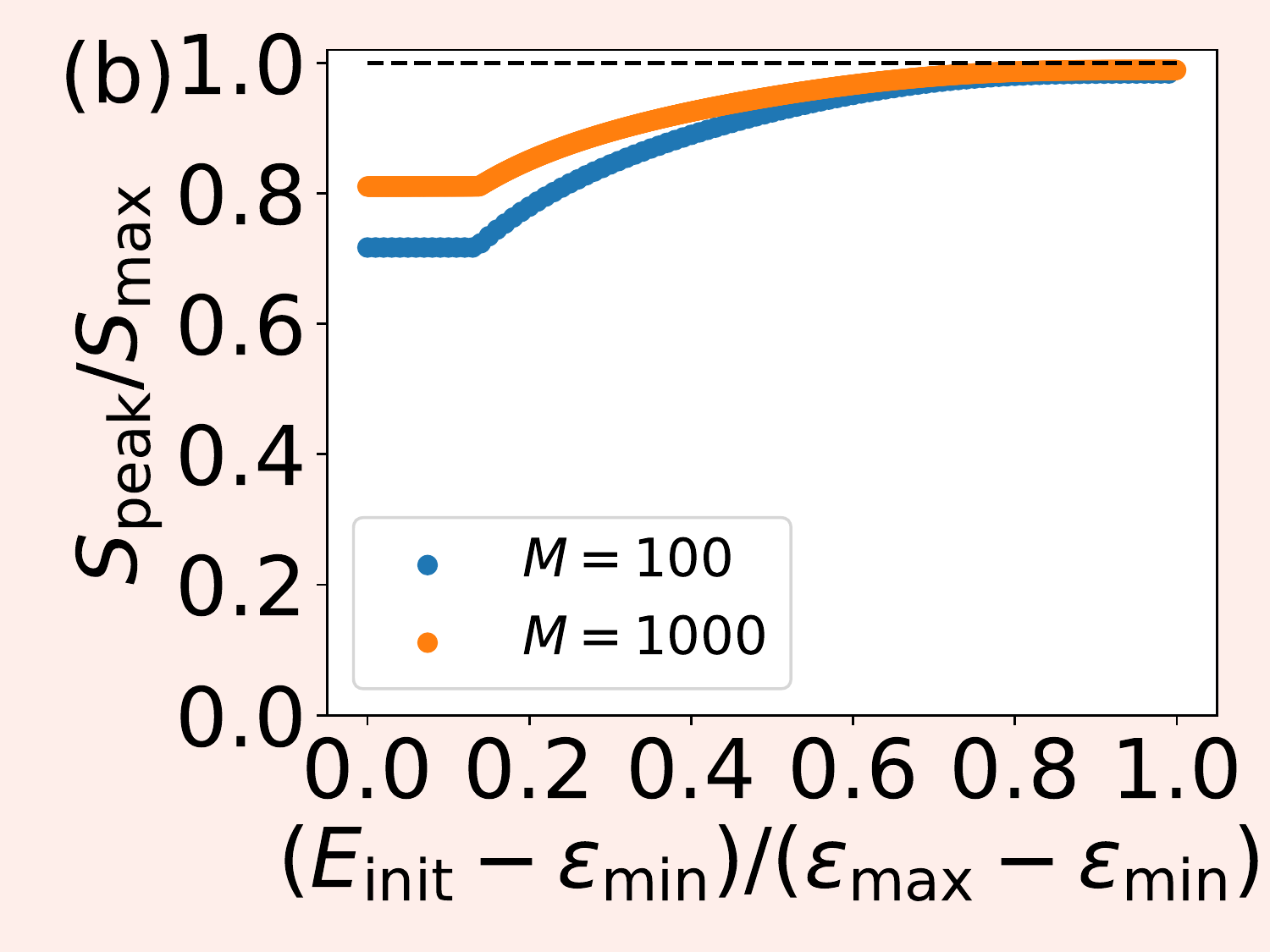}
\caption{{\bf Peak entropy versus initial state energy for the theoretical~(a) 
		 unidirectional and~(b)~bidirectional model.} The initial states are the eigenstates of the system. The initial energy is scaled between $0$, for the energy $\epsilon_{\rm min}$ of the ground state, and $1$, for the energy $\epsilon_{\rm max}$ of the most excited state.}
\label{Einit}
\end{figure}

\noindent {\bf Role of initial-state energy.}\\
Here we discuss the dependence of peak entropy on the initial-state energy. In general, a higher initial energy gives rise to a larger peak entropy.
	Figure~\ref{SvsEinit} shows the peak entropy in experiments as a function of the initial state energy. 
	One can see that initial states having energies even down to 85\% of the maximum initial energy give rise to a peak entropy close (less than 2\% distance to $S_{\mathrm{max}}$) to the maximum entropy. In Fig.~\ref{Einit}, we show the corresponding results for the theoretical models at much larger system sizes of $M=100$ and $M=1000$ states. For both models we find (almost) maximum peak entropies for sufficiently high initial energy. While for the bidirectional model, we find a large basin of initial states giving rise to maximum peak entropies, for the unidirectional model only the most excited states ensure such a large peak entropy (which is a consequence of the fact that the dynamics cannot transport probability upward in energy). However, the latter is sufficient to observe critical behavior {in time} also for the unidirectional model, as long as the most excited state can approximately be prepared experimentally. Likewise, equilibrium quantum phase transitions happen only in the ground state and not at finite temperature/energy.

\noindent {\bf Gibbs-state ansatz.}\\
We will assume that the system is described approximately by an effecitve Gibbs state in the vicinity of peak entropy. Note that this ansatz is not obvious (and cannot be justified by eigenstate thermalization as we deal with an open non-interacting system). However, it turns out to explain part of the behavior of the exact solution of the rate equation, like the observed critical exponents. Assuming the system to be described by a thermal state at effective inverse temperature $\beta$, the probability in the $m$-th eigenstate is given by
\begin{equation}\label{pm}
	p_m = \frac{e^{-\beta m \Delta}}{\sum\nolimits_m{e^{-\beta m \Delta}}}=\frac{e^{(M-m) \beta  \Delta } \left(e^{\beta  \Delta }-1\right)}{e^{M \beta  \Delta }-1}.
\end{equation}
The corresponding localization length~(participation ratio) reads
\begin{equation}
	\xi = \left(\sum\nolimits_{m} p_m^2\right)^{-1} 
	= \coth(\beta\Delta/2)\tanh(M\beta\Delta/2).
\end{equation}
In the vicinity of $\beta=0$, we obtain from a Taylor expansion
\begin{equation}
	\xi/M = 1 - \frac{1}{12}(M^2-1)(\beta\Delta)^2 + {\cal{O}}(\beta^4). 
\end{equation}
For large $M$~($M \gg 1$), one can see that $\xi$ is a function of $M\beta\Delta$.

\begin{figure}[!h]
	\centering	
	\includegraphics[width=0.99\columnwidth]{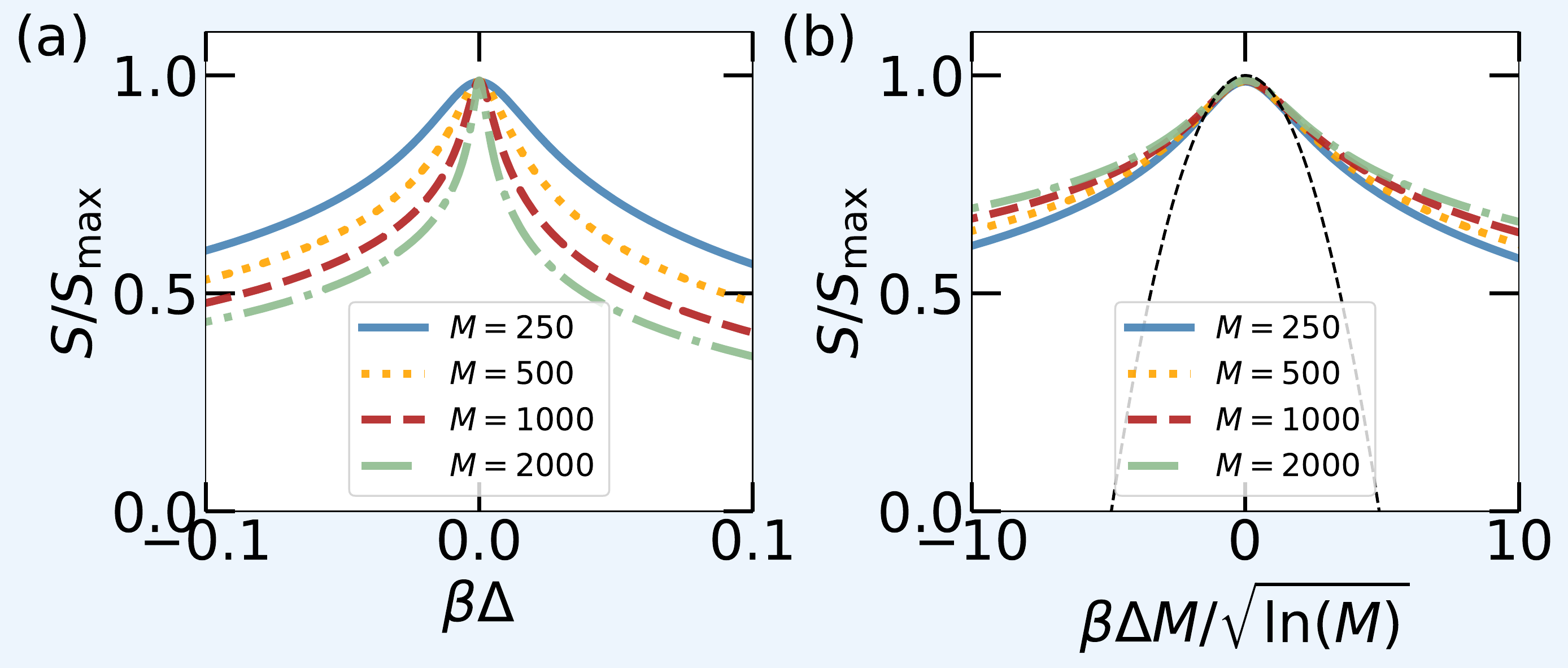}
	\includegraphics[width=0.99\columnwidth]{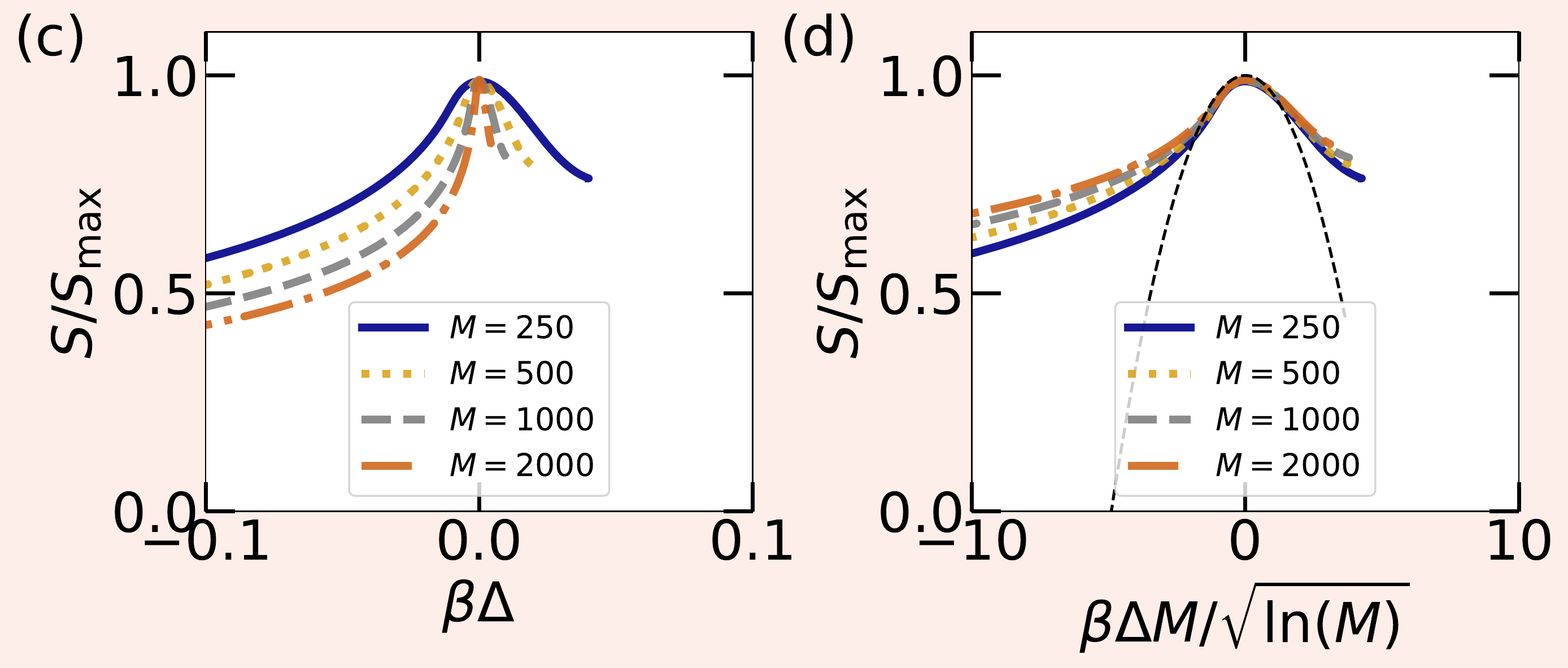}
	\caption{Entropy as a function of $\beta$ and system-size scaled $\beta$. The solid lines are the results for the unidirectional model (a,b) and for the bidirectional model (c,d). 
		The black dashed line is the result from Eq.~\eqref{Sa}.
	}\label{s_log}
\end{figure}

We can also calculate the entropy
\begin{eqnarray}\label{ss}
	S &=& -\sum\nolimits_m{p_m \log p_m} \notag\\
	&=& \beta  \Delta \left[\frac{e^{(M+1) \beta  \Delta }+M-e^{\beta  \Delta } (M+1)}{\left(e^{\beta  \Delta }-1\right) \left(e^{M \beta  \Delta }-1\right)}-M\right] \notag\\
	&&+ \ln{\frac{e^{\beta  \Delta  M}-1}{e^{\beta  \Delta }-1}}.
\end{eqnarray}
A Taylor expansion in the vicinity of $\beta=0$ then yields
\begin{eqnarray}\label{Sa}
	S/S_{\rm max} &=& 1 - \frac{1}{24}\frac{M^2-1}{\ln(M)}( \beta \Delta)^2 + {\cal O}(\beta^4)\notag\\
	&&\approx 1 - \frac{1}{24}(\frac{M}{\sqrt{\ln(M)}} \beta \Delta)^2 + {\cal O}(\beta^4) 
	\notag\\
\end{eqnarray}
It shows that the entropy is a function of $\beta \Delta M/\sqrt{\ln(M)}$ in the vicinity of $\beta=0$. In Fig.~\ref{s_log}, we plot the entropy for different system sizes as a function of $\beta\Delta$ and $\beta \Delta M/\sqrt{\ln(M)}$. Indeed, for the latter, the data are found to collapse onto each other in the vicinity of $\beta=0$.  
The result from Eq.~\eqref{Sa} is shown as the black dashed line in Fig.~\ref{s_log}, which well describes the behavior of the entropy around the peak.

\begin{figure}[!h]
	\centering	
	\includegraphics[width=0.99\columnwidth]{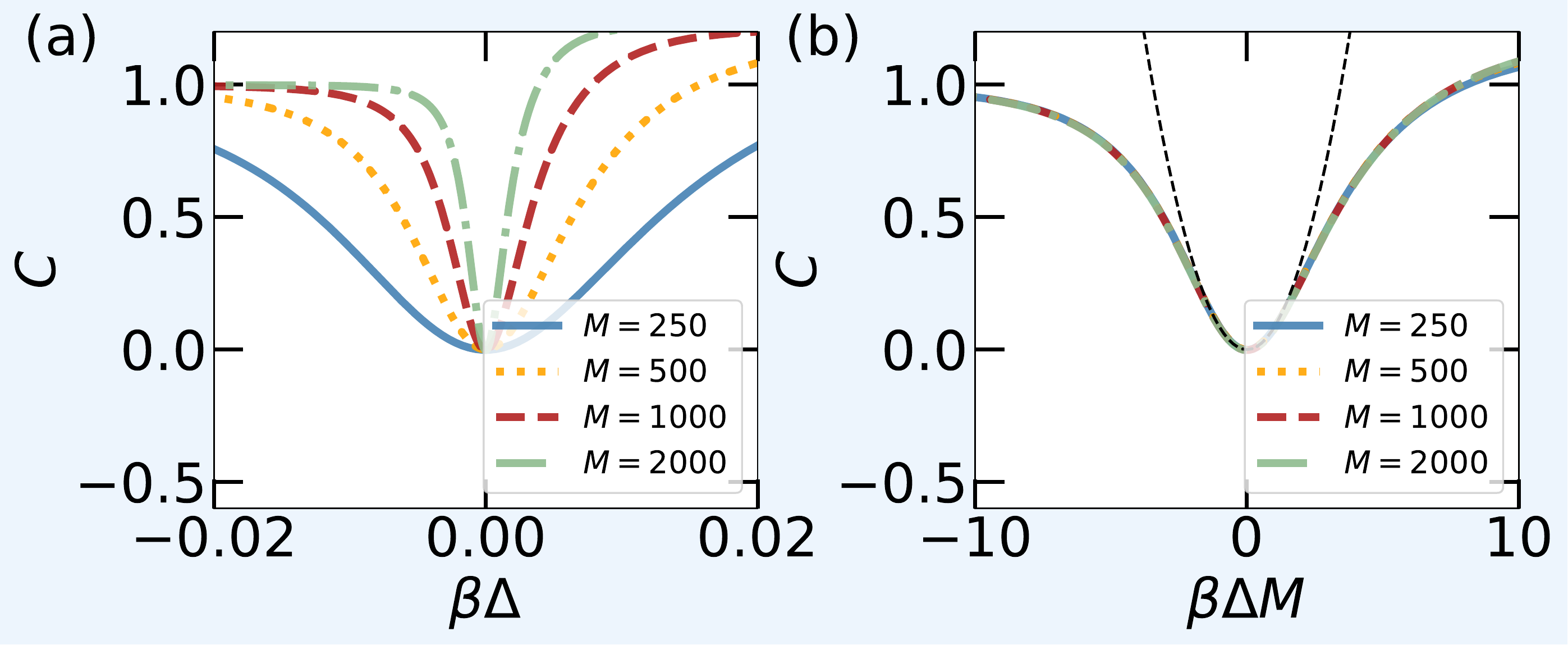}
	\includegraphics[width=0.99\columnwidth]{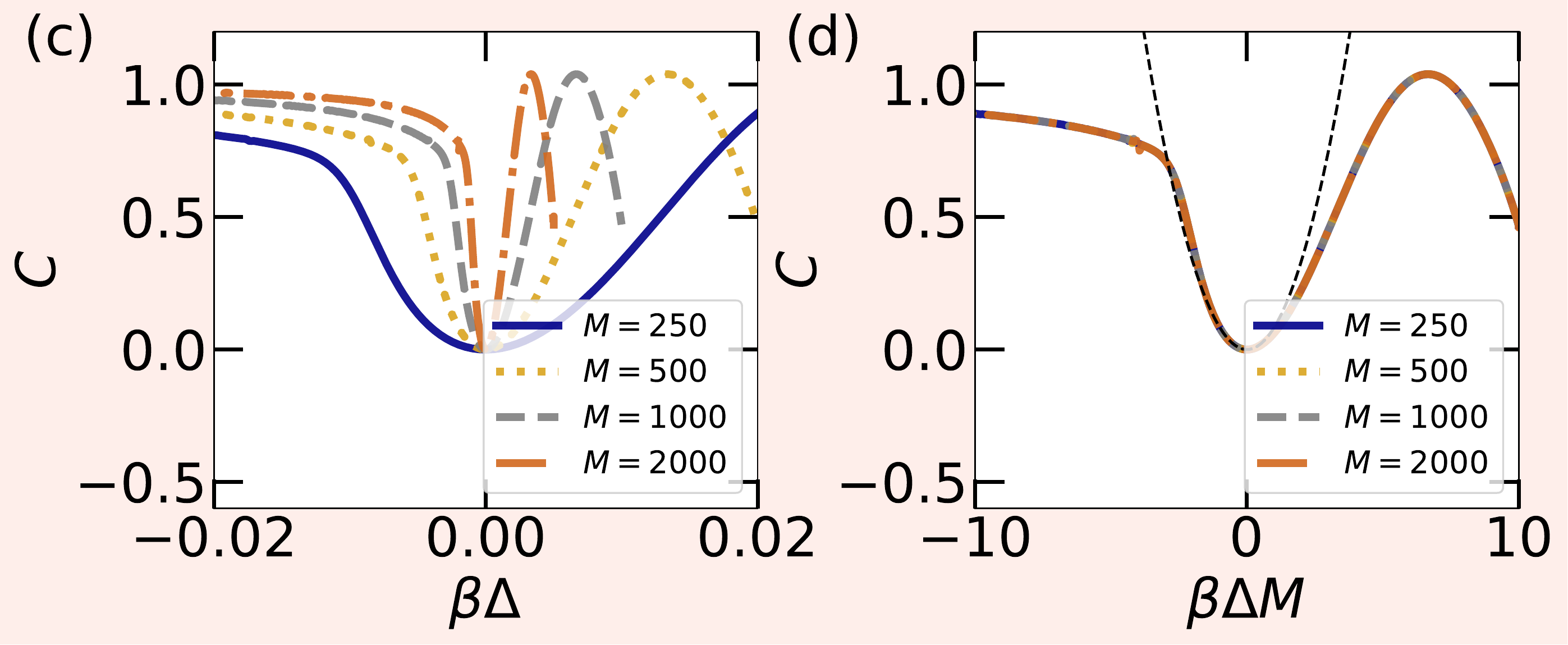}
	\caption{Effective specific heat $C \equiv d\langle H\rangle/d{T_{\rm eff}}$ as a function of $\beta$ and system-size scaled $\beta$ for the unidirectional model (a,b) and for the bidirectional model (c,d). The black dashed line is the result from Eq.~\eqref{Ca}.}\label{C}
\end{figure}

We can define an effective specific heat as 
$C \equiv d\langle H\rangle/d{T_{\rm eff}}$, with the effective temperature $T_{\rm eff} = 1/\beta$ and the mean energy $\langle H\rangle = \sum\nolimits_{m}{m\Delta p_m}$. In the vicinity of $\beta=0$, it can be shown that $C$ is given by
\begin{equation}\label{Ca}
	C = \frac{1}{12} (\beta\Delta M)^2 + {\cal O}(\beta^4).
\end{equation}
Again, we see that $C$ is a function of $\beta\Delta M$. In Fig.~\ref{C}, we show $C$ for different system sizes as a function of $\beta\Delta$ and system-size scaled $\beta$, $\beta \Delta M$. As expected, for the latter, the data collapse onto each other in the vicinity of $\beta=0$.

\noindent {\bf Continuum model.}\\

The rate equation reads
\begin{equation}
	\dot p_m = p_{m+1}R_{-,m+1} + p_{m-1}R_{+,m-1} - p_m R_{+,m} - p_m R_{-,m}.\notag
\end{equation}
By defining 
\begin{eqnarray}
	\nabla A_m &=& (A_{m+1} - A_{m-1})/2, \notag\\
	\nabla^2 A_m &=& A_{m+1} + A_{m-1} - 2A_m,
\end{eqnarray}
one can show that
\begin{eqnarray}\label{rate_eq}
	\dot p_m &=& \nabla^2\left[\bar R_m p_m\right] + \nabla [\delta R_m p_m] \notag\\
	&&+ \frac{1}{2}\nabla^2 p_m (\nabla\bar R_m +\nabla\delta R_m) + \frac{1}{2}\nabla p_m\nabla^2\delta R_m, \notag\\
\end{eqnarray}
with $$\bar R_m = (R_{+,m}+R_{-,m})/2, \quad \delta R_m = R_{-,m}-R_{+,m}.$$

Consider a system of fixed length $L$ and coordinate $x=(m/M)L\equiv m\, \delta x $. The thermodynamic limit in the discrete model $M \to \infty$ corresponds to the limit, where $x$ becomes continuous,  
\begin{equation}
	\delta x = \frac{L}{M} \to 0.
\end{equation}
The corresponding rate equation for the continuum model then reads
\begin{eqnarray}\label{ddv}
	\partial_t \rho(x,t)&=&\delta x^2\partial_x^2\left[\bar R(x)\rho(x,t)\right]\notag\\
	&&+\delta x\partial_x \left[\delta R(x)\rho(x,t)\right]+{\cal O}\left(\delta x^3\right), \notag\\
\end{eqnarray}
with $\rho(x,t)$ the probability density at $x$. For large $M$, one can neglect the high order terms ${\cal O}\left(\delta x^3\right)$, which brings Eq.~\eqref{ddv} to the drift-diffusion equation,
\begin{eqnarray}\label{dd}
	\partial_t \rho(x,t)&=&\delta x^2\partial_x^2\left[\bar R(x)\rho(x,t)\right]\notag\\
	&&+\delta x\partial_x\left[\delta R(x)\rho(x,t)\right].
\end{eqnarray}
By expanding the RHS, we obtain
\begin{eqnarray}\label{dde}
	\partial_t \rho(x,t)&=&f(x)\rho(x,t)+g(x)\partial_x\rho(x,t)\notag\\
	&&+ h(x)\partial_x^2\rho(x,t) \notag\\
	&& \equiv A(x) \rho(x,t),
\end{eqnarray}
where
\begin{eqnarray} \label{fx}
	f(x)&=&\delta x^2\partial_x^2\bar R(x)+\delta x\partial_x \delta R(x), \notag\\
	g(x) &=& 2\delta x^2\partial_x\bar R(x)+\delta x \delta R(x), \notag\\
	h(x) &=& \delta x^2 \bar R(x).
\end{eqnarray}

The formal solution of Eq.~\eqref{dde} can be written as
\begin{eqnarray}\label{rho_taylor}
	\rho(x,t) &=& e^{A(x)t}\rho(x,0) \notag\\
	&=& [1+A(x)t+A(x)^2t^2/2+\ldots]\rho(x,0), \notag\\
\end{eqnarray}
where the expansion corresponds to time-dependent perturbation theory.

Let us examine the behavior of the system starting from the maximally delocalized probability distribution,  corresponding to the maximally mixed state, which approximates the state at peak entropy. Assuming $\rho(x,0)=1/L$, corresponding to $t_\text{peak}=0$, we have
\begin{eqnarray}
	A(x)\rho(x,0) &=& f(x)/L, \notag\\
	A(x)^2 \rho(x,0) &=& u(x)/L,\notag
\end{eqnarray}
with 
\begin{equation}
	u(x)=f(x)^2+g(x)\partial_xf(x)+h(x)\partial_x^2f(x).
\end{equation}
Plugging these expressions into the expansion~\eqref{rho_taylor}, we obtain 
\begin{eqnarray}\label{rho_linear}
	\rho(x,t) =[1+f(x)t+u(x)t^2/2+{\cal{O}}(t^3)]/L,
\end{eqnarray}
In the following discussion, we will drop the terms ${\cal {O}}(t^3)$, which is valid for short positive and negative times $t$ relative to the peak time.

Plugging Eq.~\eqref{rho_linear} into the expression for the entropy 
\begin{equation}\label{entr}
	S = -\int_{0}^{L} dx \rho(x,t) \ln[\rho(x,t)\delta x],
\end{equation}
and dropping terms of ${\cal {O}}(t^3)$, we obtain
\begin{eqnarray}
	S&\simeq&\ln(M)
	\nonumber\\&&
	-\frac{1}{L}\int_{0}^{L}dx (1+ft+ut^2/2)\ln(1+ft+ut^2/2) \notag\\
	&\simeq&\ln(M)
	\nonumber\\&&
	-\frac{1}{L}\int_{0}^{L}dx (1+ft+ut^2/2)(ft+ut^2/2-f^2t^2/2)\notag\\ 
	&\simeq& \ln(M)-\frac{1}{L}\int_{0}^{L}dx {[ft+(f^2+u)t^2/2]}.
\end{eqnarray}
The normalized entropy is then approximately given by
\begin{eqnarray} 
	\frac{S}{S_{\rm max}} &\simeq& 1 - \frac{t}{L\ln(M)}\int_{0}^{L} dx f \notag\\
	&&-\frac{t^2}{2L\ln(M)}\int_{0}^{L} dx (f^2+u).
\end{eqnarray}
The integral in the linear term is approximately equal to zero according to Eq.~\eqref{rho_linear} as $\int_{0}^{L} dx \rho(x,t)=1$. Hence, the entropy reduces to
\begin{equation} \label{entropy_l}
	\frac{S}{S_{\rm max}} \approx 1  -\frac{t^2}{2L\ln(M)}\int_{0}^{L} dx (f^2+u).
\end{equation}
The energy is given by
\begin{equation}
	E = \frac{\Delta}{\delta x}\int_0^L{dx x\rho(x,t)}.
\end{equation}
Plugging Eq.~\eqref{rho_linear} into the above expression, one obtains in leading order
\begin{equation}
	E \simeq \frac{1}{2}\Delta M + t\frac{\Delta  M}{L^2}\int_0^L{dx x f}.
\end{equation}
The effective inverse temperature is given by
\begin{equation}\label{beta}
	\beta = \frac{dS/dt}{dE/dt} \approx - t\frac{\int_{0}^{L} dx (f^2+u) L}{\int_0^L{dx x f \Delta M}}.
\end{equation}
In terms of $\beta$, the entropy reads
\begin{equation}\label{beta}
	S/S_{\rm max} \approx 1 - \frac{(\beta\Delta M)^2}{\ln(M)} \frac{(\int_{0}^{L} dx x f)^2}{2L^3 \int_{0}^{L} dx (f^2+u)}
\end{equation}
and one can see that $$1-S/S_{\rm max} \propto \frac{(\beta \Delta M)^2}{\ln(M)}.$$ 
This is in consistent with the analytic results from Gibbs state assumption as described in the previous section.

By plugging Eq.~\eqref{rho_linear} into the definition for the localization length, 
\begin{equation}
	\eta = \xi \delta x = \left[\int_0^L {dx \rho(x,t)^2}\right]^{-1},
\end{equation}
we find
\begin{equation}
	\eta^{-1} \approx \frac{1}{L} + \frac{1}{L^2}\int_0^L {dx (f^2+u) t^2}.
\end{equation}
In the finite-size scaling analysis for the unidirectional model with time as control parameter, the overall strength of the bath-induced rates scales with the system size as $M^{1.16}$, in which case the entropy reaches peak around the same time for different system sizes. Therefore, $\sqrt{f^2+u}  \propto \delta x M^{1.16} \propto M^{0.16}$. It indicates that the localization length is a function of $M^{0.16}t$, which is consistent with the numerical results as shown in Fig.~\ref{tFSS}.

In terms of $\beta$, the localization length reads
\begin{eqnarray}
	\eta/L &=& \xi/M \notag\\
	&\approx& \left\{1+(\beta\Delta M)^2 \frac{(\int_{0}^{L} dx x f)^2}{ L^3\int_{0}^{L} dx (f^2+u)}\right\}^{-1} \notag\\
	&\approx& 1-(\beta\Delta M)^2  \frac{(\int_{0}^{L} dx x f)^2}{ L^3\int_{0}^{L} dx (f^2+u)}.
\end{eqnarray}
Hence, $\xi/M$ is a function of $\beta\Delta M$.
Similar analysis shows that the effective specific heat is also a function of $\beta\Delta M$.

\end{document}